\documentclass[twocolumn,epjc3]{svjour3} 
\usepackage{graphicx}
\usepackage{amsmath}
\usepackage{hyperref}
\usepackage{xspace}
\usepackage[utf8]{inputenc}
\usepackage{color}
\usepackage{todonotes}
\usepackage{lipsum}
\usepackage{cite}
\usepackage[english]{babel}

\def\beq{\begin{equation}}
\def\beqn{\begin{eqnarray}}
\def\eeq{\end{equation}}
\def\eeqn{\end{eqnarray}}

\newcommand\GeV{\ensuremath{\mathrm{GeV}}\xspace}
\newcommand\TeV{\ensuremath{\mathrm{TeV}}\xspace}

\newcommand\mfl{\ensuremath{m_{{\mathrm4}\ell}}\xspace}
\newcommand\mtltn{\ensuremath{m_{{\mathrm2}\ell{\mathrm2}\nu}}\xspace}
\newcommand\mll{\ensuremath{m_{\ell\ell}}\xspace}
\newcommand\kt{k_{\scriptscriptstyle\rm T}}

\newcommand\HWpp{{\tt HERWIG++}\xspace}
\newcommand\Herwigpp\HWpp
\newcommand\HERWIGPP\HWpp

\newcommand\PYTHIAn{{\tt PYTHIA~\!\!8}\xspace}

\newcommand\POWHEG{{\tt POWHEG}}
\newcommand\POWHEGBOX{{\tt POWHEG-BOX}\xspace}

\newcommand\RES{{\tt POWHEG-BOX-RES}\xspace}
\newcommand\OL{{\tt OpenLoops}\xspace}
\newcommand\ggfl{{\tt gg4l}\xspace}

\newcommand\FASTJET{{\tt Fastjet}\xspace}

\newcommand\back{bkgd\xspace}
\newcommand\interf{intf\xspace}
\newcommand\signal{signal\xspace}

\newcommand{\Asigl}{A_{\rm signal}}
\newcommand{\Abkgd}{A_{\rm bkgd}}

\journalname{Eur. Phys. J. C}

\begin{document}

\title{Four-lepton production in gluon fusion at NLO matched to parton showers}

\author{
Simone~Alioli\thanksref{addr1,e1}
\and
Silvia~Ferrario Ravasio\thanksref{addr2,addr3,e2}
\and
Jonas~M.~Lindert\thanksref{addr4,e3}
\and
Raoul~R\"ontsch\thanksref{addr5,e4}
}

\thankstext{e1}{e-mail: simone.alioli@unimib.it}
\thankstext{e2}{e-mail: silvia.ferrarioravasio@physics.ox.ac.uk}
\thankstext{e3}{e-mail: j.lindert@sussex.ac.uk}
\thankstext{e4}{e-mail: raoul.rontsch@cern.ch}

\institute{Universit\`a degli Studi di Milano-Bicocca \& INFN, Piazza della Scienza 3, Milano 20126, Italia\label{addr1}
\and
Rudolf Peierls Centre for Theoretical Physics, University of Oxford, Parks Road, Oxford OX1 3PU, UK\label{addr2}
\and
IPPP, Department of Physics, Durham University, Durham DH1 3LE, UK\label{addr3}
\and
Department of Physics and Astronomy, University of Sussex, Brighton BN1 9QH, UK\label{addr4}
\and
Theoretical Physics Department, CERN, 1211 Geneva 23, Switzerland\label{addr5}
}

\date{Preprint: IPPP/20/79, CERN-TH-2021-022 \\\\ Received: date / Accepted: date  ******** \ \today \ *********  }

\maketitle

\abstract{
We present a calculation of the next-to-\linebreak leading order (NLO) QCD
corrections to gluon-induced electroweak gauge boson pair production, $gg \to ZZ$ and $gg \to W^+W^-$, matched to  the \PYTHIAn{} parton shower in the \POWHEG{} approach. The calculation consistently incorporates the continuum background, the Higgs-mediated $gg\to H^* \to VV$ process, and their interference. We consider leptonic decay modes of the massive vector bosons and retain offshell and non-resonant contributions.  The  processes considered are loop-induced at leading order and thus contain two-loop virtual contributions as well as loop-squared real contributions. Parton-shower effects are found to be marginal in inclusive observables and quite sizeable in observables that are exclusive in additional jet radiation. The Monte Carlo generator presented here allows for realistic experimental effects to be incorporated in state-of-the-art precision analyses of diboson production and of the Higgs boson in the offshell regime.
}


\section{Introduction}
\label{sec:intro}	
	
One of the main objectives of Run 3 of the Large Hadron Collider (LHC) will be the further investigation of the Higgs sector.
Most studies directly targeting the Higgs boson will focus on its onshell production and subsequent decay. 
Indeed, one might naively expect that the cross section to produce an offshell Higgs boson is negligible, due to the extremely narrow width of the Higgs boson of about 4 MeV in the Standard Model (SM).
However, contrary to this expectation, it is known that approximately 10\% of $gg \to H^* \to VV$ events are produced with an invariant mass $m_{VV}$ above the $2 m_V$ production threshold~\cite{Kauer:2012hd}.
The importance of the offshell region for Higgs phenomenology was further highlighted in Ref.~\cite{Caola:2013yja}, 
which showed that a comparison of onshell and offshell data can provide stringent constraints on the width of the Higgs boson (see also Refs.~\cite{Campbell:2013una,Campbell:2013wga}). 
While later work indicated that such constraints are not model-independent, 
they also revealed the potential of using offshell data to probe the couplings of the Higgs boson~\cite{Anderson:2013afp,Gainer:2013rxa,Englert:2014ffa,Logan:2014ppa,Gainer:2014hha,Englert:2014aca,Azatov:2014jga}.
Offshell analyses have been performed by both ATLAS~\cite{Aad:2015xua,Aaboud:2018puo} and CMS~\cite{Khachatryan:2014iha,Khachatryan:2015mma,Khachatryan:2016ctc,Sirunyan:2019twz}, 
and have succeeded in constraining the Higgs boson width to $\mathcal{O}$(10\,MeV). This is several orders of magnitude smaller than a direct constraint, which is limited by the detector resolution. 
Nevertheless, offshell analyses are currently still limited by the available statistics.
Further studies of offshell Higgs boson production will therefore be a key component of the investigations of the Higgs sector during both Run 3 and in the high luminosity phase of the LHC.

In this paper, we will focus on the production of an offshell Higgs boson through gluon fusion and its subsequent decay into a pair of  electroweak gauge bosons. To this end we consider the signal Higgs production  process $gg \to H^* \to VV$
together with the corresponding continuum background process $gg \to VV$	and their interference.
We study the two diboson modes $VV=\{ZZ,W^+W^-\}$ and we assume leptonic decays of the diboson pair.
In the following, for brevity, we often denote the processes according to the intermediate diboson resonances ($ZZ$,~  $W^+W^-$). However by this we always refer to the full four-lepton offshell processes, including the interference between $Z$ and offshell photon production.

The signal process proceeds predominantly through a top-quark loop.
For onshell Higgs production, the top-quark mass is the largest scale in the process and can be approximated as infinitely heavy, allowing this loop-induced process to be reduced to a tree-level one.
Using this approximation, the next-to-next-to-next-to-leading order (N3LO) corrections to Higgs production have been computed~\cite{Anastasiou:2015vya,Anastasiou:2016cez,Mistlberger:2018etf}.
However, this approximation is not valid for offshell Higgs production, since the virtuality of the Higgs may be comparable to (or even larger than) the top-quark mass.
This means that  a leading-order (LO) prediction for offshell Higgs production requires the computation of a one-loop amplitude with the full top mass dependence, while the next-to-leading order (NLO) correction requires a two-loop amplitude.
By itself, this would not be so onerous, but there is a second reason why predictions for offshell Higgs production are more demanding than for onshell Higgs production.
It is well-known that the interference effects between the signal $gg \to H^* \to VV$ and the background process $gg \to VV$ can be sizeable and thus must be taken into account~\cite{Kauer:2012hd}.
Moreover, as we discussed above, the impact of top quarks in the loops cannot be neglected, and this means that in the computation of the background amplitudes $gg \to VV$ the contribution from both massless and massive quarks circulating in the loops should be considered.

Results for offshell Higgs production including the mass dependence of quarks in the loop and interference effects are  known at LO ~\cite{Binoth:2008pr,Kauer:2012hd,Campbell:2013una,Campbell:2013wga}.
Results in the presence of an additional radiated jet have also been presented~\cite{Cascioli:2013gfa,Campbell:2014gua}.	
At NLO, the two-loop $gg \to VV$ amplitudes for massless quarks circulating in the loop have been known for several years~\cite{Caola:2015ila,vonManteuffel:2015msa}.
However, the corresponding amplitudes for massive quark loops have only recently become available~\cite{Agarwal:2020dye,Bronnum-Hansen:2020mzk, Bronnum-Hansen:2021olh}.
This means that a fully consistent NLO prediction with the exact dependence on the top-quark mass for offshell Higgs production is in sight but still not available.

However, NLO calculations including interference effects
have been presented based on an expansion in $1/m_t$~\cite{Melnikov:2015laa,Campbell:2016ivq,Caola:2016trd}.
This expansion is not valid for high energies, but has been shown to work well below the top-pair production threshold $2m_t$.
In fact, Ref.~\cite{Campbell:2016ivq} uses a conformal mapping and Pad\'e approximants to extend the results beyond the top-pair threshold.
More recently, it has been demonstrated that using an expansion in $1/m_t$  together with a threshold expansion as inputs for Pad\'e approximants can lead to
improved estimates for both $gg \to HH$ and $gg \to VV$ amplitudes~\cite{Grober:2017uho,Grober:2019kuf}. In Ref.~\cite{Davies:2020lpf} the massive two-loop amplitude for $gg \to  ZZ$ has been computed in the high-energy expansion $s,|t| \gg m_t^2$, which opens the door for a NLO description of this process in the phase space \mbox{$m_{VV} > 2 m_t$}.
However, even disregarding these methods, there is a significant region of the offshell phase space with \mbox{$ m_{VV} < 2\,m_t$}  
in which the $1/m_t$ expansion is expected to be reliable, and hence where a good approximation to the NLO corrections can be obtained. 
We base the Monte Carlo generator presented here on such an approximation, following the calculation of Ref.~\cite{Caola:2016trd}.
In the future, the generator can easily be extended to also cover the region  $m_{VV} > 2\,m_t$
by replacing the massive two-loop amplitudes. 

Reliable NLO corrections to the continuum background $gg \to VV$ alone can be obtained ignoring heavy quark contributions (or these can be incorporated via a reweighting of the massless two-loop amplitude with the LO mass dependence). They are available in the literature both for $gg \to ZZ$~\cite{Caola:2015psa,Grazzini:2018owa} and $gg \to W^+W^-$~\cite{Caola:2015rqy,Grazzini:2020stb}.~\footnote{The results of Refs.~\cite{Grazzini:2018owa,Grazzini:2020stb} also include the offshell Higgs contribution, however without investigating it explicitly.
} Formally these are of $\mathcal{O}(\alpha_S^3)$
with respect to the LO $pp\to VV$ process, i.e. they contribute beyond the order of the known NNLO corrections to the quark-induced channels~\cite{Cascioli:2014yka,Gehrmann:2014fva,Grazzini:2015hta,Grazzini:2016ctr,Heinrich:2017bvg} - yet they yield phenomenologically relevant contributions.

The NLO results of Refs.~\cite{Caola:2015psa,Caola:2015rqy,Campbell:2016ivq,Caola:2016trd,Grazzini:2018owa,Grazzini:2020stb} are at fixed-order parton level, meaning that they do not account for radiation beyond one additional jet.
This, together with the fact that unweighted events are not available, hinders the use of these calculations in experimental analyses.
In this paper, we report on NLO calculations for offshell Higgs production, including interference effects, matched to parton showers using the \POWHEG{} method~\cite{Nason:2004rx,Frixione:2007vw,Alioli:2010xd,Jezo:2015aia}. The implementation extends earlier work by two of us~\cite{Alioli:2016xab} that considered the background process $gg \to ZZ \to 4\ell$ only.
Furthermore, in contrast to Ref.~\cite{Alioli:2016xab}, here we also include the contribution from $qg$- and $q\bar{q}$-initiated channels. 
This implementation allows the generation of unweighted events with additional radiation included through the parton shower, and should facilitate the use of the NLO calculations in experimental analyses. The corresponding \RES generator \ggfl will be made publicly available in due time.

The paper is organized as follows. 
In Sec.~\ref{sec:computational_setup}, we briefly discuss the technical details involved in the \linebreak parton-level calculation as well as in the matching procedure.
In Sec.~\ref{sec:setup}, we summarize the numerical inputs that we use.
In Sec.~\ref{sec:fixedorder}, we present fixed-order results validating our calculation and investigate the applied approximations. Finally in 
Sec.~\ref{sec:nlopsresults} we present numerical results 
for  $ZZ$ and $WW$ production matched to parton showers. 
We conclude in Sec.~\ref{sec:conclusions_outlook}.

\section{Computational setup} 
\label{sec:computational_setup}

In this section, we describe the matching of the NLO calculation of gluon-induced four-lepton production to parton showers through the \POWHEG{} method implemented in \RES. We first describe  the structure of the fixed-order NLO computation and then discuss several details relevant for the matching to \PYTHIAn{}.

\subsection{Structure of the NLO computation}
\label{sec:computation}

We begin by summarizing the salient features of the NLO calculation, and refer the reader to Ref.~\cite{Caola:2016trd} for additional discussion.
As mentioned in the previous section, we need to consider both Higgs-mediated amplitudes $gg \to H^* \to VV$ as well as continuum production $gg \to VV$ amplitudes. We therefore write the full amplitude for gluon-induced $VV$ production as
\begin{equation}
A = \Asigl + \Abkgd
\end{equation}
where $\Asigl$ refers to Higgs-mediated amplitudes, while $\Abkgd$ refers to amplitudes without any Higgs propagators. 
Squaring this equation gives
\begin{equation}
|A|^2 = |\Asigl|^2 + |\Abkgd|^2 + 2{\rm Re}\left(\Asigl \Abkgd^*\right)\,.
\end{equation}
Upon integrating over the phase space for the final state particles, the first two terms on the right-hand side give the signal and background results, respectively, while the third term gives the interference contribution
\begin{equation}
  \mathrm{d}\sigma_{\rm full} = \mathrm{d}\sigma_{\rm signal} + \mathrm{d}\sigma_{\rm bkgd} + \mathrm{d}\sigma_{\rm intf}.
\end{equation}
In Secs.~\ref{sec:fixedorder} and ~\ref{sec:nlopsresults} we will present results for these contributions separately, as well as for their sum $\mathrm{d} \sigma_{\rm full}.$

As mentioned in the previous section, the LO amplitudes for both $\Asigl$ and $\Abkgd$ are well known~\cite{Glover:1988rg,Matsuura:1991pj,Zecher:1994kb,Binoth:2008pr,Kauer:2012hd,Campbell:2013una,Campbell:2013wga}.
At NLO, we have to compute the real and virtual corrections to  $\Asigl$ and $\Abkgd$. 
The corrections to $\Asigl$ have been known for some time~\cite{Ellis:1987xu,Spira:1995rr,Harlander:2005rq,Aglietti:2006tp}. 
On the other hand, the NLO corrections to the background amplitude $\Abkgd$ are more involved, and deserve a separate discussion.

We begin by examining the virtual corrections to the $gg \to ZZ$ process. In this case, one can clearly separate massless loops of the first five flavours, and massive top-quark loops.
The virtual (two-loop) amplitudes for the former are known~\cite{vonManteuffel:2015msa,Caola:2015ila}, and  we construct these using the {\tt ggVVamp} library~\cite{vonManteuffel:2015msa}. 
Results for two-loop amplitudes with massive quarks were presented very recently~\cite{Agarwal:2020dye, Bronnum-Hansen:2021olh}. However, here we follow the approach of Refs.~\cite{Melnikov:2015laa,Caola:2016trd} and use an expansion in $1/m_t$ for the massive amplitudes.
This implies that our NLO results for the $ZZ$ production process are only valid below the top pair production threshold $m_{ZZ} < 2m_t$. 
Finally, we need to include double-triangle amplitudes, where each triangle can have either massless or massive quarks in the loop.
We employ analytic results for these amplitudes taken from Refs.~\cite{Hagiwara:1990dx,Campbell:2007ev}.

\begin{figure}[t!]
  \begin{center}
  	  \begin{tabular}{cccc}
  	  	  \includegraphics[scale=0.21]{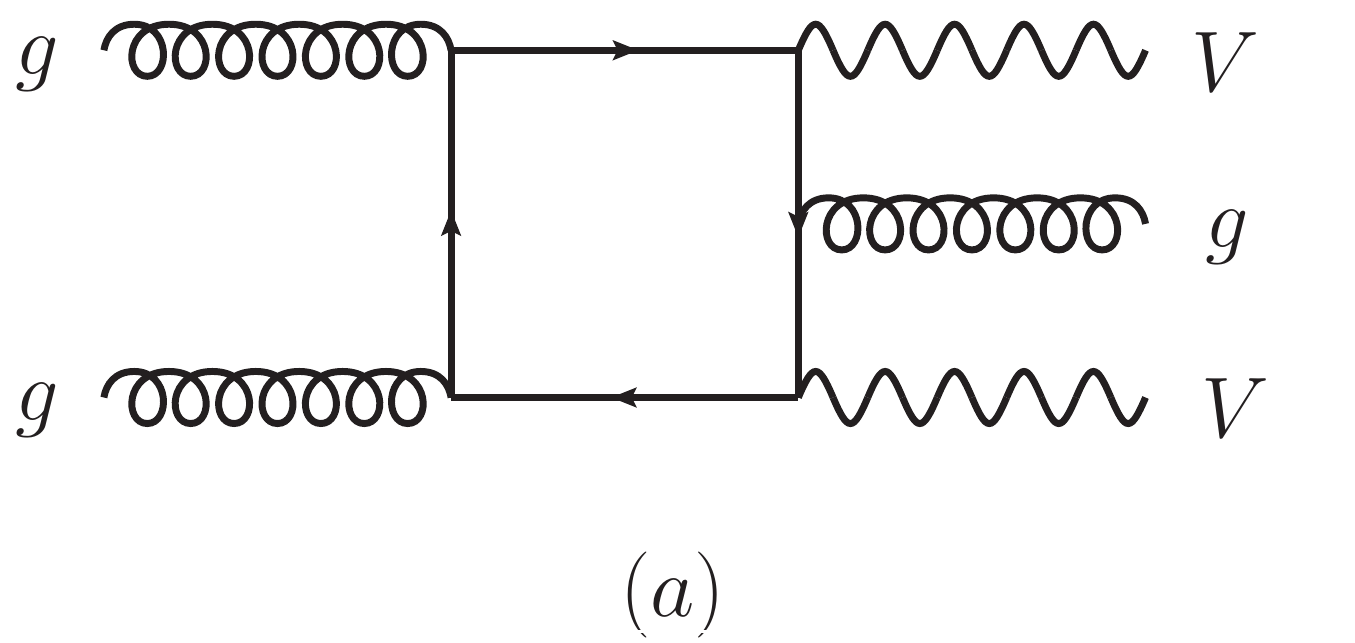} & & 
  \includegraphics[scale=0.21]{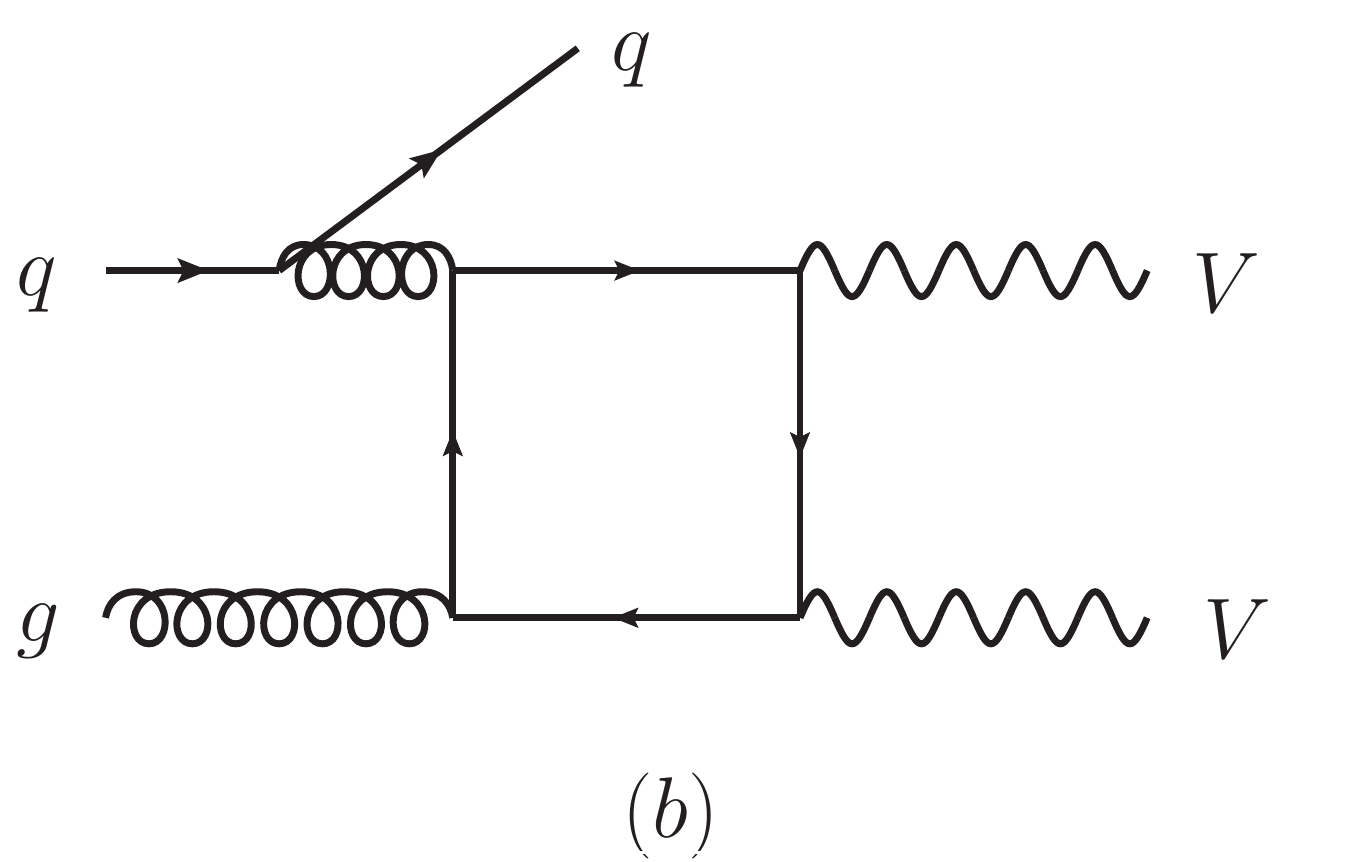}\\[0ex]
      \includegraphics[scale=0.21]{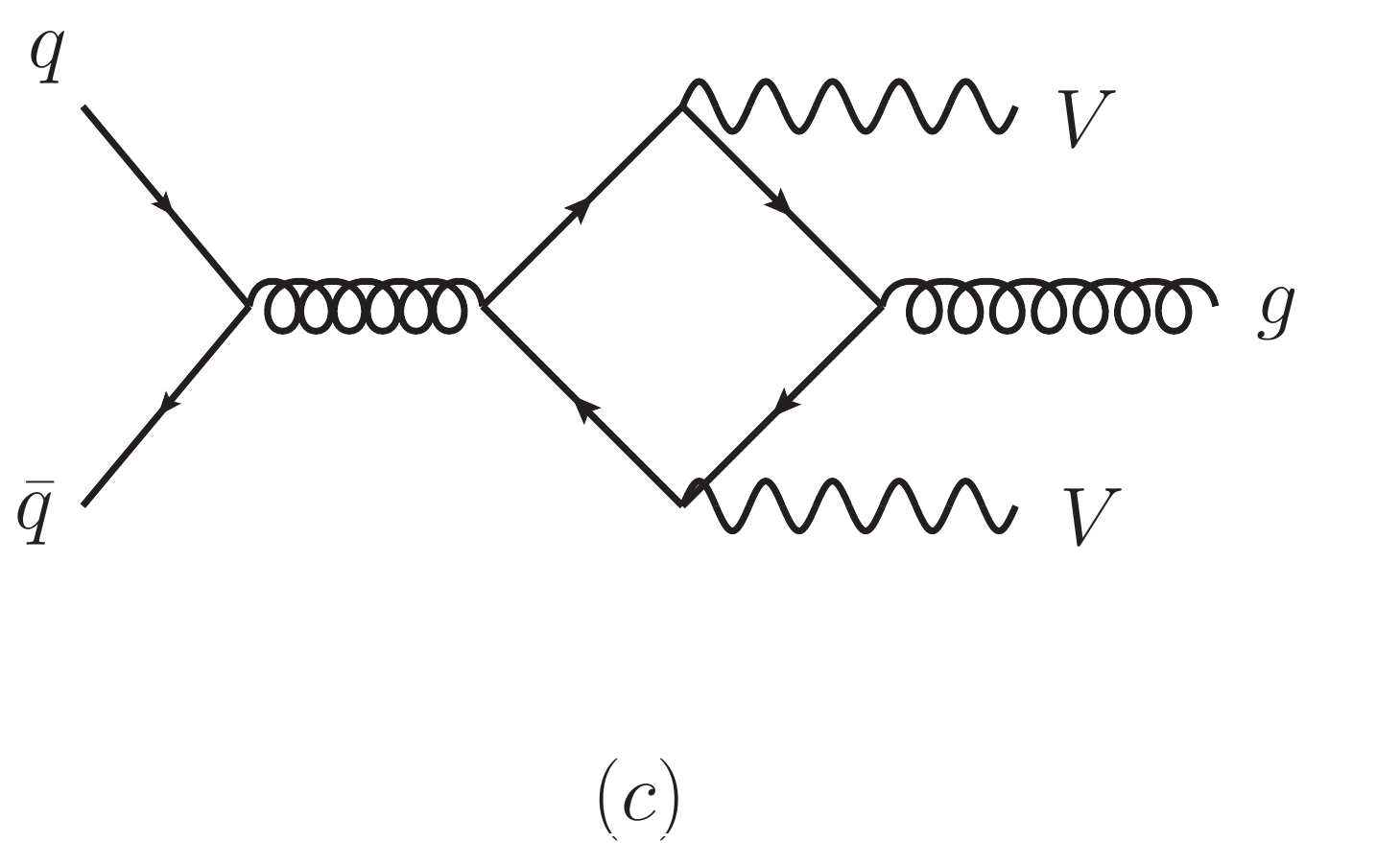} & &
   \includegraphics[scale=0.21]{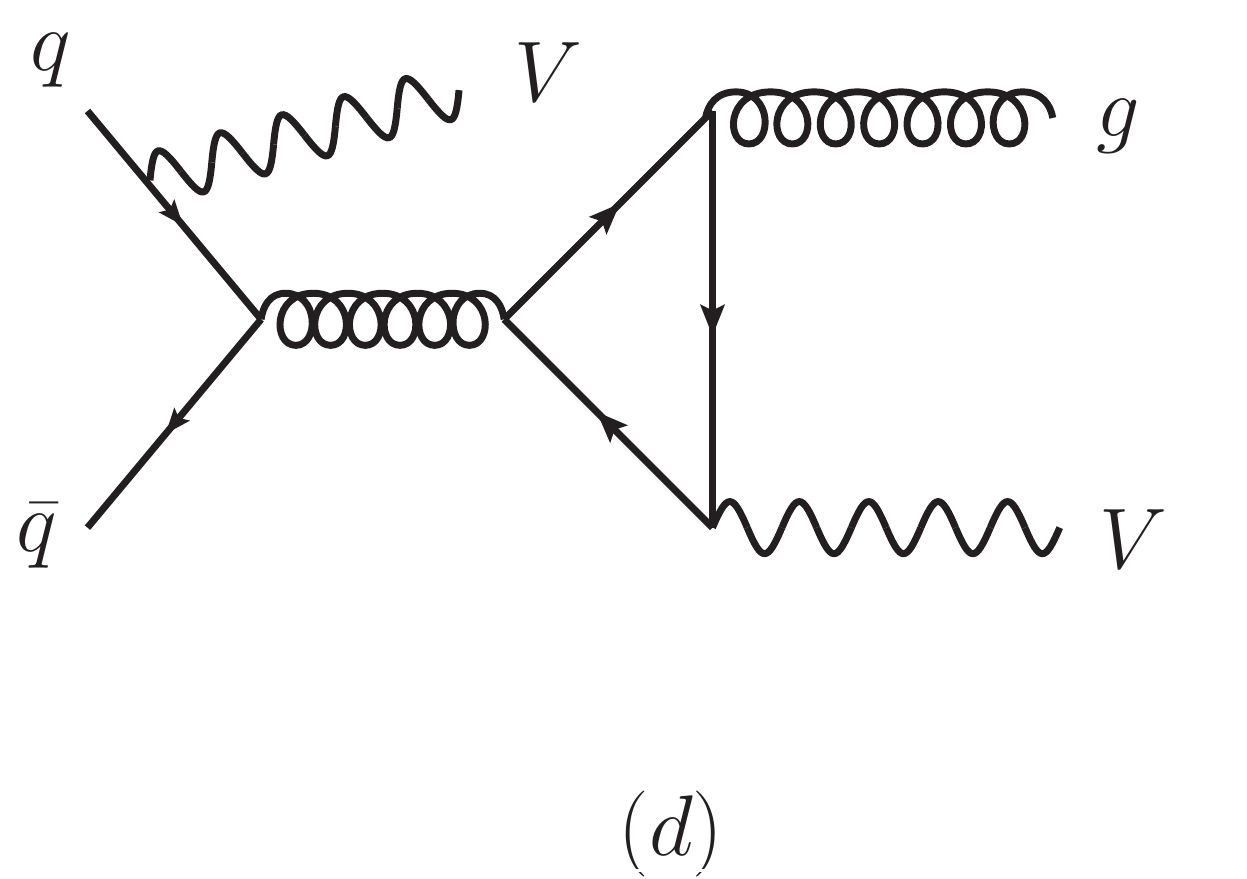}\\[0ex]
  	  \end{tabular}
 \end{center}
   \caption{\label{fig:diagrams}%
     Real corrections to gluon-induced $VV$ production, with different partonic channels.}
\end{figure}

We now discuss the case of $WW$ production.
Since top and bottom quarks mix in the loop, there is no longer a clear division into massive and massless loops.
For this reason, Ref.~\cite{Caola:2016trd} only considered four massless quark flavours in the loop for $WW$, neglecting the third generation entirely.
Here we take a slightly different approach.
We compute the two-loop amplitudes $\Abkgd^{\rm 2loop}$ using {\tt ggVVamp} assuming massless quarks for four active flavours. We then reweight the amplitudes as follows
\begin{align}
  \label{eq:rwgt}
\Abkgd^{\rm 2loop, rwgt} = \Abkgd^{\rm 2loop} (u,d,s,c)\frac{|\Abkgd^{\rm 1loop} (u,d,s,c,t,b)|^2}{|\Abkgd^{\rm 1loop} (u,d,s,c)|^2}\, 
\end{align}
where $\Abkgd^{\rm 1loop} (u,d,s,c,t,b)$ is the one-loop amplitude with full mass dependence for the third-generation quarks and  $\Abkgd^{\rm 1loop} (u,d,s,c)$ is the one-loop amplitude with four active flavours.
\footnote{We note that results for the two-loop $gg \to WW$ amplitudes with massive top quarks  in the loop were recently presented in Ref.~\cite{Bronnum-Hansen:2020mzk}.}
We will comment on the accuracy of this approach in Sec.~\ref{sec:masseffects}.

The real corrections to $gg \to VV$ include both the purely gluonic channel $gg \to VV + g$ as well as channels with initial state quarks
 $qg \to VV+q$ and $q\bar{q} \to VV+g $ (see Fig.~\ref{fig:diagrams}) and their crossings.
At $\mathcal{O}(\alpha_s^3)$, the former can be  unambiguously identified as corrections to the loop-induced process $gg \to VV$.
The $qg$ and $q\bar{q}$ channels are more intricate.
These channels also appear in the $\mathcal{O}(\alpha_s^3)$ corrections to the $q\bar{q} \to VV$ process,
and it is not possible to parametrically distinguish these corrections from the corrections to the loop-induced process that we are interested in.
For this reason, these channels were not included in Ref.~\cite{Caola:2016trd}.
On the other hand, there is no obstacle to computing these corrections,
as they form a gauge invariant subset and their only infrared singularities are removed through the collinear renormalization of the parton distribution functions.
Indeed, results for these channels were included in Refs.~\cite{Grazzini:2018owa,Grazzini:2020stb}.
In this paper, we choose to include these channels in our nominal predictions at NLO and also in the results after matching to the  parton shower.
At the same time we will also investigate the impact of these channels in Sec.~\ref{sec:qgchannel}, so that both their magnitude and their impact on the scale variation can be properly assessed.
Note that we define these contributions to include any amplitudes with \textit{at least one} vector boson attached to a closed fermion loop.
In particular, amplitudes such as those represented in Fig.~\ref{fig:diagrams}(d) (contributing to $gg\to ZZ$) are included.
\footnote{In our code, the user can choose to switch off amplitudes with one vector boson attached to an external quark line using the {\tt ol\_noexternalvqq} flag.
  The user can also turn off the $qg$ and $q\bar{q}$ channels altogether with the {\tt select\_real} flag.}
  
We compute all the real correction loop-squared amplitudes using {\tt OpenLoops 2}~\cite{Cascioli:2011va,Buccioni:2019sur} including massless and massive quark contributions in the loop, allowing us to retain the full dependence on the top quark mass (and bottom quark mass where applicable).
This is in contrast to the approach of Ref.~\cite{Caola:2016trd} where the real amplitudes for $gg \to ZZ+g$ involving a top-quark loop were computed using an expansion in $1/m_t$.
Finally, we note that our calculation includes single-resonant amplitudes in all partonic channels.

In order to ensure numerical stability across the whole phase-space, including in the IR regions close to the soft or collinear limits of the real radiation, we rely on the {\tt OpenLoops} stability system, which automatically reevaluates all phase-space points with the two reduction methods implemented in {\tt Collier}~\cite{Denner:2016kdg}. For unstable points matrix elements are set to zero. We verified that varying the corresponding threshold \texttt{stability\_kill2} by a factor of 10 around a central value of $10^{-2}$ leaves all results unchanged (see Ref.~\cite{Buccioni:2019sur} for documentation of this threshold parameter). 

In order to optimize the treatment of all the colorless resonances, we take advantage of the \RES framework~\cite{Jezo:2015aia}, which, despite being  specifically designed to handle the subtractions when intermediate colored resonances are present, can at the same time improve the phase-space sampling of any resonance. This is achieved by first manually specifying the resonance histories.\footnote{The automatic resonance-finding algorithm in \RES is not yet able to handle resonances in loop-induced contributions.} The \RES then decomposes the cross section into contributions associated to a well-defined resonance structure, which are enhanced on that particular cascade chain. Each contribution is  separately integrated at this point with a dedicated resonance-aware phase space sampling which makes use of a resonance-aware subtraction procedure. The resonance-aware subtraction makes use of a mapping from a real to the underlying Born configuration that preserves the virtuality of intermediate resonances.
Due to the absence of QCD divergences in the resonances, the resonance-aware subtraction is strictly speaking not necessary for the processes considered here. However, we choose to adopt it because its usage improves the statistical errors for observables directly probing the resonance structure.
The last essential feature of the \RES implementation is the ability to generate remnants and regular events even when the corresponding cross section is negative, which was not possible in previous versions of the \POWHEGBOX that were instead assuming them to be positive. Despite usually being squares of matrix elements, in this process remnants and regulars contributions might indeed assume negative values in the calculation the interference terms.
Technical details about necessary modifications in \RES to deal with the processes at hand are given in \ref{sec:mod}.

\subsection{Matching to \PYTHIAn{}}
\label{sec:matching}

We  next discuss the matching of the NLO calculation of $gg \to VV$  to the \PYTHIAn{} parton shower in the framework of \RES.

The resonance structure that we construct at the partonic level is further preserved by the parton shower by specifying the input resonance cascade chain at the Les Houches event level (LHE) and making sure that the shower does not distort it through recoil effects.
This is achieved by using the {\tt PowhegHooks} class in \PYTHIAn{}. However, the {\tt PowhegHooks} class needs to know the number of final state particles involved in the LO process once the resonance decays are stripped. Since in the \RES{} this number is not fixed,  we modified the {\tt PowhegHooks} class accordingly, following the recipe adopted in Ref.~\cite{FerrarioRavasio:2019vmq}.

The \PYTHIAn{} parton shower implements two recoil schemes for initial-state radiation~(ISR): in both cases the recoil is always applied to all final-state particles to absorb the transverse momentum imbalance due to ISR off an initial-initial dipole. In the default scheme~\cite{Sjostrand:2004ef} the same is also done for initial-final dipoles. There is, however, also the option to use a fully local scheme~\cite{Cabouat:2017rzi}, in which when an initial-state emission takes place from an initial-final dipole, the final-state spectator absorbs the transverse-momentum recoil and the other particles in the event are left unchanged.

The default recoil scheme is the recommended option to handle the $s$-channel production of colour singlets, while the alternative one was originally designed to handle deep inelastic scattering and vector boson fusion events.
Since at LO the process considered in this paper describes the production of a colour singlet, we maintain as our baseline the default recoil scheme of \PYTHIAn{}.
However, since in principle we can also generate events with a hard final state jet, 
in our numerical results we compare the two recoil prescriptions for exclusive observables that might be sensitive to this choice.

In all our showered predictions we include underlying event simulation and hadronization effects. However, in order to simplify the identification of the leptons, we turn off QED radiation and the decay of unstable hadrons.

\section{Numerical setup}
\label{sec:setup}

In this section we present the numerical inputs for the results presented in the following sections.

Coupling and mass input parameters are fixed to the following values:
\begin{align*}
m_{\mathrm{Z}}&=91.1876~\GeV\;, & \Gamma_{\mathrm{Z}}&=2.4952~\GeV\;,\\
m_{\mathrm{W}}&=80.3980~\GeV\;, & \Gamma_{\mathrm{W}}&=2.1054~\GeV\;,\\
m_{\mathrm{H}}&=125.1~\GeV\;, & \Gamma_{\mathrm{H}}&=4.03\cdot10^{-3}~\GeV\;,\\
m_{\mathrm{t}}&=173.2~\GeV\;, & G_{F}&=1.16639 \cdot 10^{-5}\,{\rm GeV}^{-2}\,,
\end{align*}
where $G_{F}$ denotes the Fermi constant and 
\begin{align*}
\alpha=\frac{\sqrt{2}}{\pi}m_W \sin^2\theta_W\,,
\end{align*}
with the real-valued weak mixing angle
\begin{align*}
\sin^2\theta_W=1-\frac{m_W^2}{m_Z^2}\,.
\end{align*}

In general, the \ggfl generator allows for finite bottom-quark masses. In this work, we mostly use the $N_F=5$ flavour scheme and treat the bottom quark as massless $m_{\mathrm{b}}=0~\GeV$.  Only in 
Sec.~\ref{sec:validation}, where we validate against the results of Ref.~\cite{Caola:2016trd}, do we use a non-zero bottom-quark mass, and there we choose $m_{\mathrm{b}}=4.5~\GeV$. 

We use the partonic luminosities and strong coupling from the \texttt{NNPDF30\_lo\_as\_0130} and the \linebreak \texttt{NNPDF30\_nlo\_as\_0118}  sets~\cite{Ball:2014uwa} for the validation against the results of Ref.~\cite{Caola:2016trd} that we present in Sec.~\ref{sec:validation}. For all other results, we use the \texttt{NNPDF31\_nlo\_as\_0118} set~\cite{Ball:2017nwa}.

We consider center-of-mass energies of 13~\TeV, and set as renormalization and factorization scales for all modes
\begin{align}
\mu=\mu_{R}=\mu_{F}&=\frac{\mfl}{2}\,,
\end{align}
where 
\begin{align}
\mfl^{2}=\left(\sum_{i \in \{\ell, \nu\}} p_{i}\right)^{2}\,.
\end{align}
We obtain scale uncertainty bands by independently varying  the renormalization and factorization scales by a factor of two and omitting antipodal variations.

At the generator level the following kinematic cuts are applied in the $ZZ$ channel,
\begin{align}
5~\GeV&<\mll<180~\GeV,\label{eq:cut_mll}\\
70~\GeV&<\mfl<340~\GeV\label{eq:cut_m4l}\,.
\end{align}
We need to impose such an upper cut on $\mfl$ because, as discussed in the previous sections,  the virtual corrections are computed using a $1/m_t$ expansion which is no longer valid for large values of $\mfl$~\cite{Caola:2016trd}.
For $WW$ production we only require
\begin{equation}
 \mtltn >1~\GeV,
\end{equation}
to ensure the renormalization and factorization scales remain inside the perturbative domain. We do not impose any transverse momentum or rapidity requirements on the final-state leptons.

In order to avoid numerical instabilities of the loop-induced amplitudes we need to impose additional mild technical cuts at the generation level. For the Born kinematics, we discard configurations where the transverse momentum of the vector boson is smaller than 100~MeV.
For the real corrections, we neglect configurations where the transverse momentum of the radiated parton is smaller than 100~MeV, as also done in Ref.~\cite{Alioli:2016xab}. Indeed this region only gives rise to power-suppressed contributions that do not significantly change the total cross-section. We verified that our results are independent of these technical cuts varying them by a factor of 5 from 0.1 GeV to 0.5 GeV.

Finally, we reconstruct jets with the anti-$\kt$
algorithm~\cite{Cacciari:2008gp} as implemented in the \FASTJET{}
package~\cite{Cacciari:2005hq, Cacciari:2011ma}, with jet radius
$R=0.4$ and $p_{T,j} > 20$\,GeV.

\section{Fixed-order NLO results}
\label{sec:fixedorder}

In the following we present selected fixed-order results that we used to validate our
implementation and to investigate the accuracy of the applied approximation for the 
treatment of mass effects in the virtual corrections.

\subsection{Validation}
\label{sec:validation}

\begin{table}[t!]
\centering
\small
\begin{tabular}{c| c | c || c | c }
\multicolumn{5}{c}{ $ZZ$: $gg \to  e^{+} e^{-} \mu^{+} \mu^{-}$}\\
\hline

  &\multicolumn{2}{|c||}{ \RES}
  &\multicolumn{2}{|c}{  Ref.~\cite{Caola:2016trd} }     \\
  \hline
contrib
 &LO  [fb]
 &NLO  [fb]
 &LO  [fb]
 &NLO  [fb]\phantom{\Big|} \\
  \hline
\back   & 2.898(1)  & 4.482(6)  & 2.90(1)  & 4.49(1)  \\
\signal & 0.0431(1) & 0.0745(2) & 0.043(1) & 0.074(1) \\
\interf &-0.1542(3) &-0.2870(4) &-0.154(1) &-0.287(1) \\
\end{tabular}
\qquad 
\begin{tabular}{c| c | c || c | c }
\hline
\multicolumn{5}{c}{ $W^+W^-$: $gg \to  e^{+} \nu_e \mu^{-} \bar\nu_{\mu}$}\\
\hline
  &\multicolumn{2}{|c||}{ \RES}
  &\multicolumn{2}{|c}{  Ref.~\cite{Caola:2016trd} }     \\
  \hline
contrib
 &LO  [fb]
 &NLO  [fb]
 &LO  [fb]
 &NLO  [fb]\phantom{\Big|} \\
  \hline
 \back         &  48.92(6)  &74.62(7)   & 49.0(1)  &74.7(1)\\
 \signal       &  48.24(8)  &83.31(5)   &48.3(1)   &83.35(2)\\
 \interf       &  -2.24(1)  &-4.20(2)   &-2.24(1)  &-4.15(1)\\
\end{tabular}
\caption{Comparison of LO and NLO cross sections for the signal, background, and interference contributions to $gg \to ZZ\to e^{+} e^{-} \mu^{+} \mu^{-}$ (top) and $gg \to W^+W^-\to e^{+} \nu_e \mu^{-} \bar\nu_{\mu}$ (bottom) with those of Ref.~\cite{Caola:2016trd}. The $qg$- or $q\bar q$-induced channels are not considered.}
\label{table:total_XS}
\end{table}

As a validation of our implementation, we compare the fixed-order LO and NLO cross sections for $gg \to ZZ\to e^{+} e^{-} \mu^{+} \mu^{-}$ and 
 $gg \to W^+W^-\to e^{+} \nu_e \mu^{-} \bar\nu_{\mu}$ against the results of 
Ref.~\cite{Caola:2016trd} in Tab.~\ref{table:total_XS}, with selection cuts as specified in Ref.~\cite{Caola:2016trd}.
The signal, background and interference contributions are shown  separately.  
 Following the approach of Ref.~\cite{Caola:2016trd}, a finite bottom-quark mass $m_b$ is used everywhere in the signal ($|\Asigl|^2$).
 For the $ZZ$ channel, the background ($|\Abkgd|^2$) is computed with $m_b=0$ and the interference ($2 \rm{Re}(\Asigl\Abkgd^*)$) is computed with a finite $m_b$ besides from the virtual contribution which is computed analytically in a mixed-mass scheme, where the bottom mass is neglected in the background amplitude $\Abkgd$.\footnote{Conversely to the calculation in Ref.~\cite{Caola:2016trd}, we cannot easily use a different bottom mass value for $\Abkgd$ and $\Asigl$ when evaluating the interference using the Born and real matrix elements provided by \OL.}
 For the $WW$ channel, $\Abkgd$ is evaluated with $n_f=4$, for both the background and the interference channels.

For this validation we do not consider quark-initiated channels in the real contribution, consistent with Ref.~\cite{Caola:2016trd}.
Within numerical accuracy we find convincing agreement between the two calculations at LO and NLO.

\begin{figure}[tb]
\includegraphics[width=0.48\textwidth]{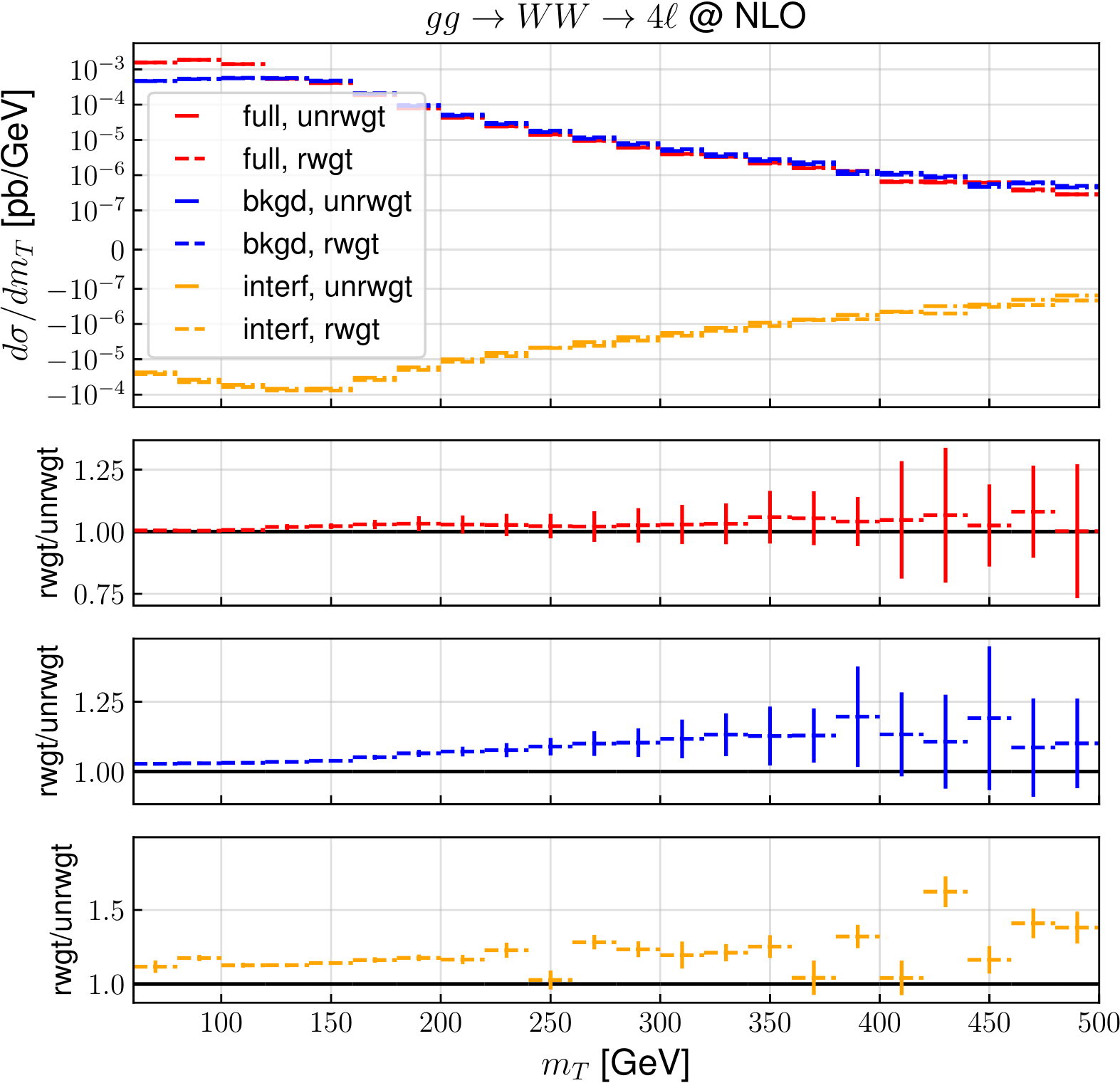}

\caption{Differential distribution in the transverse mass $m_{T}$ of the four lepton system in $gg \to e^{+} \nu_e \mu^{-} \bar\nu_{\mu} $ at fixed-order NLO. We show
	results with and without reweighting of the heavy-quark mass effects in the virtual amplitude using dashed and dashed-dotted curves, respectively. The comparison is shown for the full (red), the background (blue) and the interference (orange) contributions. The lower panes show the bin-by-bin ratios of the results without reweighting  to those with reweighting.
	For the nominal prediction we use a symlog scale with a linear threshold=$10^{-7}$.
}
\label{fig:WW-NLO-rwgt}
\end{figure} 

\begin{figure}[tb]
\includegraphics[width=0.48\textwidth]{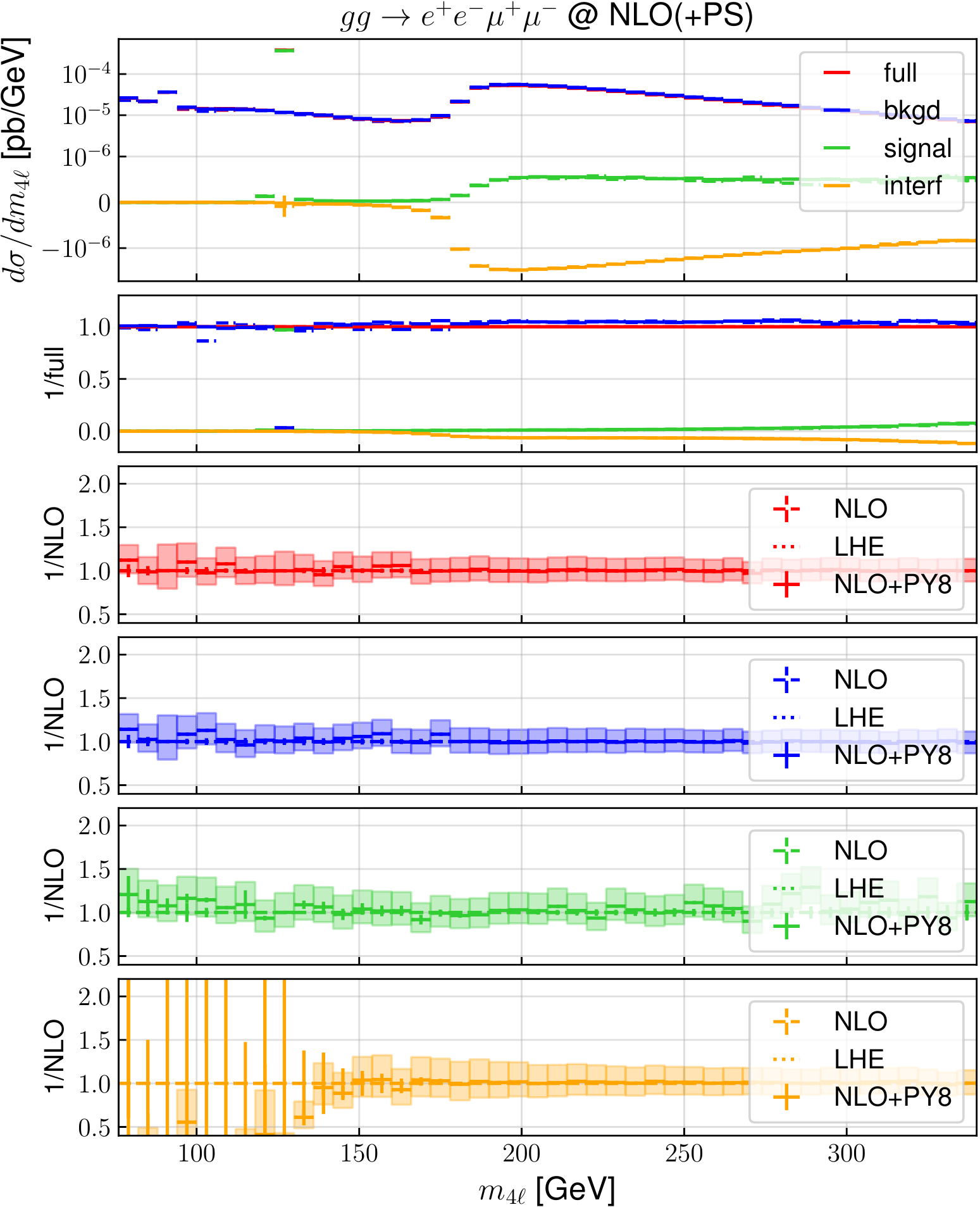}
\caption{Differential distribution in invariant mass $m_{4\ell}$ of the four-lepton system in $gg \to e^{+} e^{-} \mu^{+} \mu^{-}$ at NLO matched to \PYTHIAn{}. The upper panel shows nominal predictions at fixed-order NLO (dashed) for the background (blue), the signal (green) and the interference (orange) separately and their sum (red) together with NLO+PS predictions (solid). For the nominal prediction we use a symlog scale with a linear threshold=$10^{-6}$.
  The first ratio plot shows the relative yield of the different contributions with respect to the full, both at the fixed-order NLO level and also after parton shower. The lower four ratio plots show the LHE level (dotted) and fully showered corrections with respect to fixed-order NLO for the sum of all contributions (second ratio plot), the background only (third ratio plot),  the signal only (fourth ratio plot), and for the interference only (fifth ratio plot). The band associated to NLO+PS predictions indicates the $7$-point scale variation uncertainty.}
\label{fig:ZZ-ZZmass}
\end{figure} 
\begin{figure}[tb]
\includegraphics[width=0.48\textwidth]{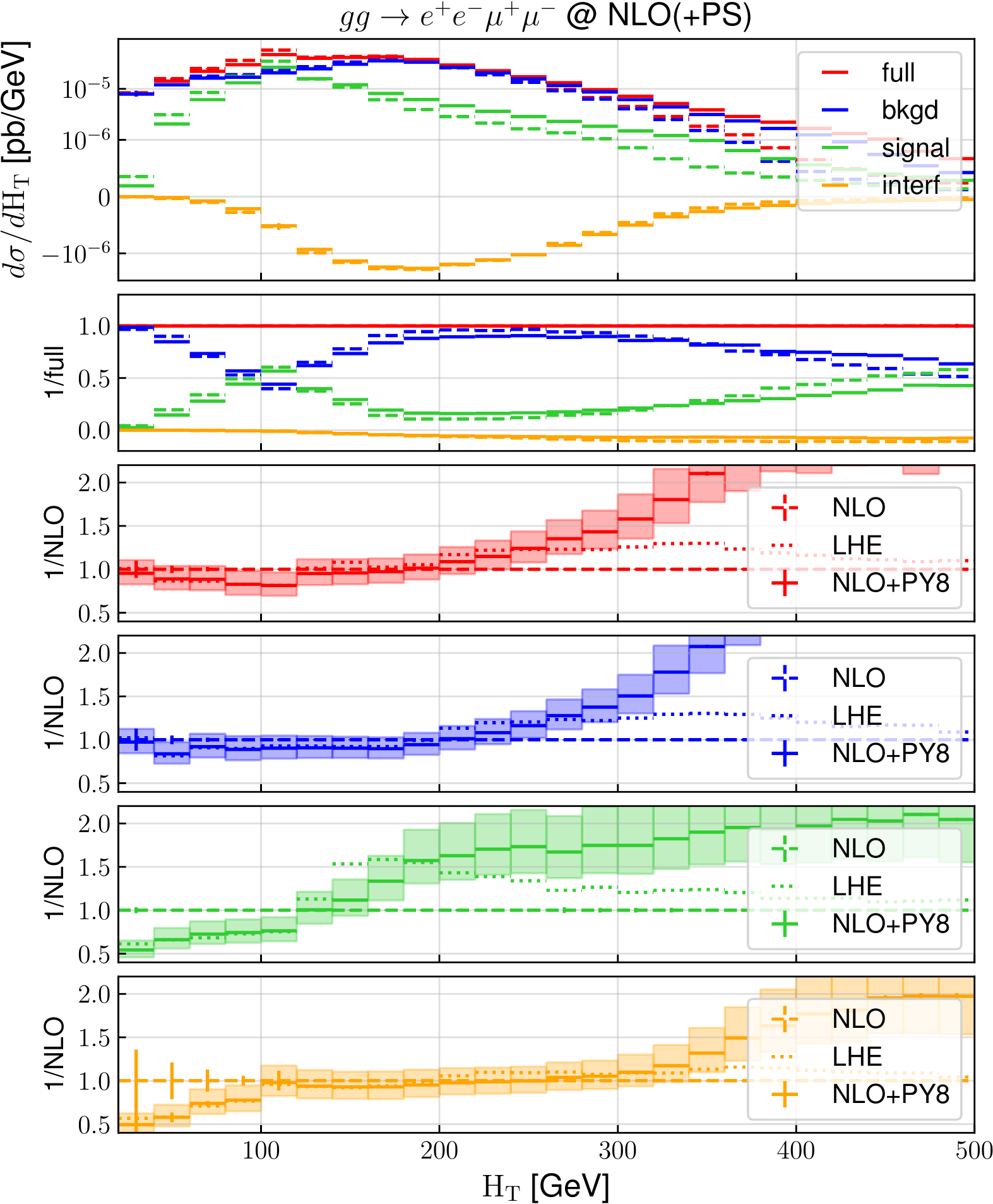}
\caption{Differential distribution in $H_{\rm T}$ in $gg \to e^{+} e^{-} \mu^{+} \mu^{-}$ at NLO matched to \PYTHIAn{}. Predictions, colour coding and bands as in Fig.~\ref{fig:ZZ-ZZmass}.}
\label{fig:ZZ-HTtot}
\end{figure} 

\subsection{Mass effects}
\label{sec:masseffects}

As discussed in Sec.~\ref{sec:computation}, the massive contributions to the two-loop virtual amplitudes are incorporated via approximations in our calculation. All other ingredients including the real amplitudes retain full 
mass dependence. For the $ZZ$ process the approximation for the massive 
two-loop amplitudes is based on an expansion in $1/m_t$. As discussed in detail in Ref.~\cite{Caola:2016trd} the resulting accuracy is estimated to be at the percent level for $m_{ZZ} < 2m_t$ and quickly deteriorates beyond this. 
For the $WW$ process we  reweight the massless two-loop amplitudes
as detailed in  Eq.~\ref{eq:rwgt}. In order to gauge the impact of this
reweighting and thus the accuracy of the applied approximation, we compare against results obtained  using only two massless generations for two-loop $\Abkgd$ amplitudes,  while all other contributions are computed using full mass dependence as usual.
We plot these results for the transverse mass $m_{T}$ of the four lepton system in $W^+W^-$ production in Fig.~\ref{fig:WW-NLO-rwgt}. This observable is defined as
\begin{align}
\label{eq:mt}
	m_{T} = \sqrt{2\, E_{\rm T,miss}\, p_{\rm T,\ell^+\ell^-} \,(1-\cos
	(\phi_{\rm miss, \ell^+\ell^-})}
	 \,,
\end{align}
where the missing transverse energy $E_{\rm T,miss}$ is given by the neutrino momenta at truth level and
$\phi_{\rm miss, \ell^+\ell^-}$ is the angle between the sum of the neutrino momenta and the sum of the lepton momenta.
For the background contribution the effect of the reweighting is at the few percent
level for the bulk of the cross section and increases up to about $15\%$-$20\%$ at large transverse masses. For the interference the impact is at the 
$15\%$ level inclusively and  mildly increases in the tail of the  transverse mass
distribution. The sum of all contributions --  which also includes the signal where no
approximations are needed -- receives inclusive variations due to the reweighting procedure of $7\%$. In the tail this increases to $10\%$-$15\%$.

\section{NLO results matched to parton showers}
\label{sec:nlopsresults}

In this section we present our numerical results matched to the \PYTHIAn{} parton shower. We consider the different-flavour decay modes $gg \to e^{+} e^{-} \mu^{+} \mu^{-}$ and \linebreak
 $gg \to e^+ \nu_e \mu^{-} \bar\nu_{\mu}$ and for simplicity denote them $ZZ$ and $W^+W^-$ production respectively. The same-flavour leptonic decay modes will also be made available in the \ggfl generator, but are not the focus of this study.

\begin{figure}[tb]
\includegraphics[width=0.48\textwidth]{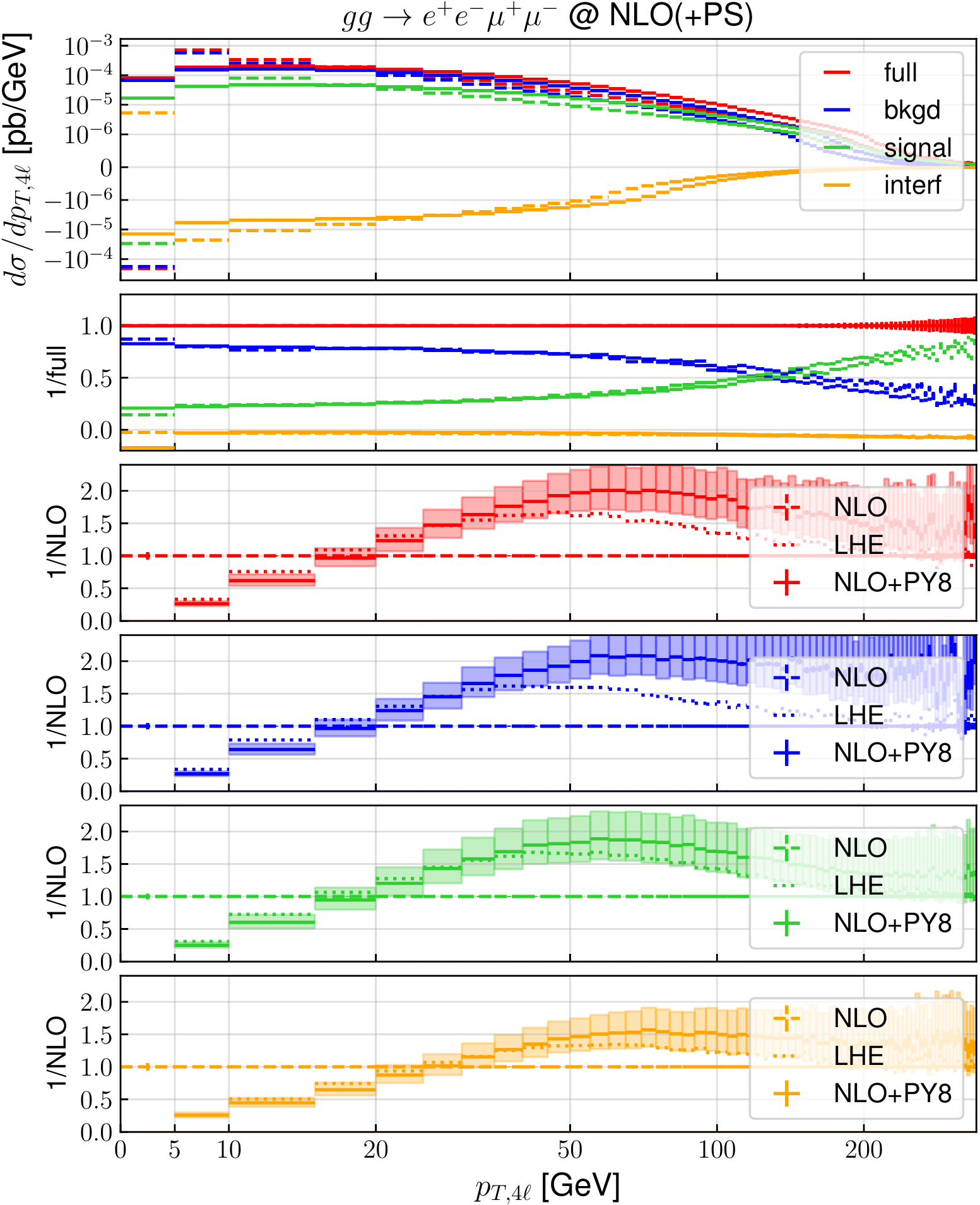}
\caption{Differential distribution in the transverse momentum of the four lepton system $p_{\rm T,4\ell}$ in $gg \to e^{+} e^{-} \mu^{+} \mu^{-}$ matched to \PYTHIAn{}. Predictions, colour coding and bands as in Fig.~\ref{fig:ZZ-ZZmass}.
}
\label{fig:ZZ-pTZZ}
\end{figure} 

\begin{figure}[tb]
\includegraphics[width=0.48\textwidth]{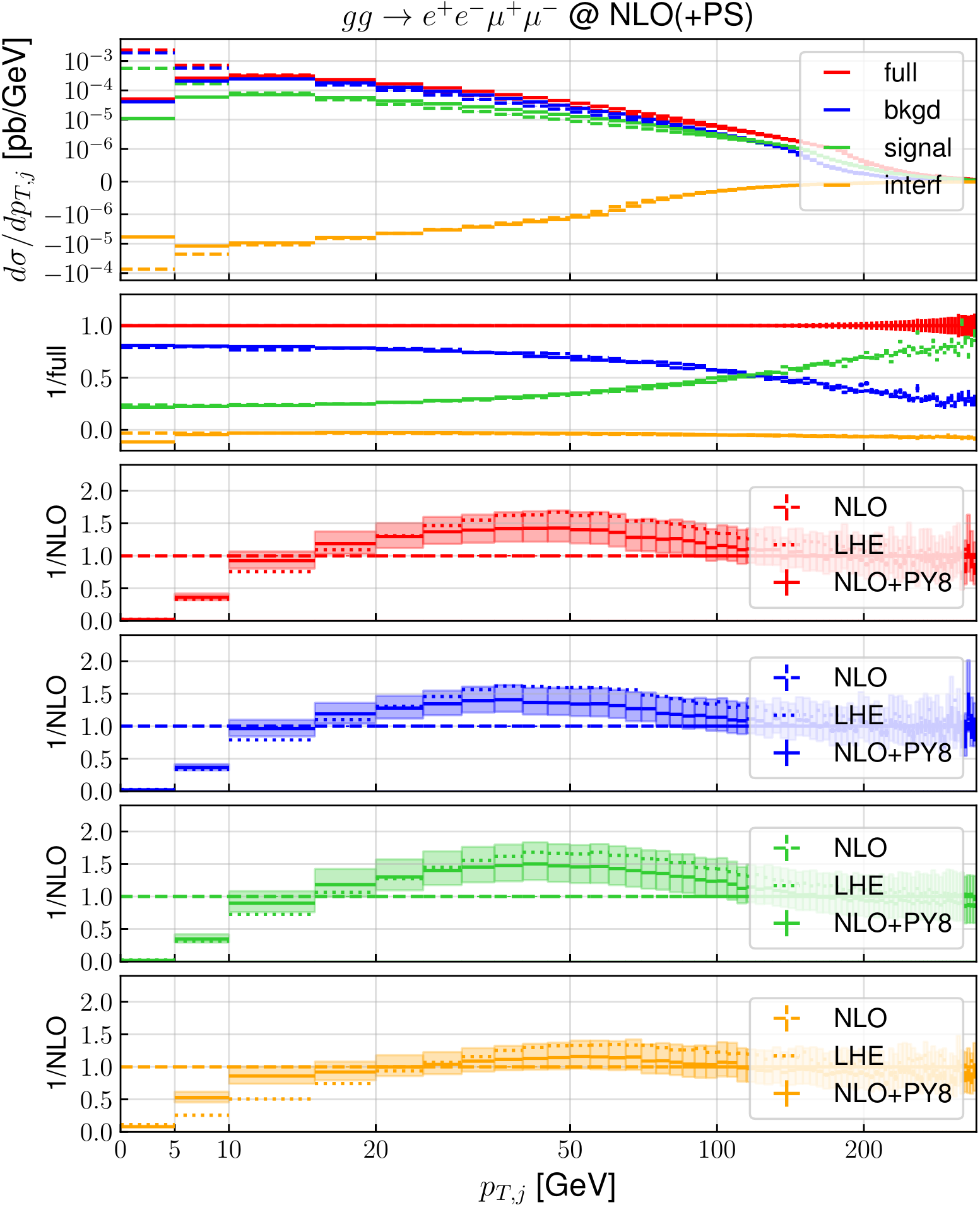}
\caption{Differential distribution in the transverse momentum of the hardest jet $p_{\rm T,j_1}$ in $gg \to e^{+} e^{-} \mu^{+} \mu^{-}$ at NLO matched to \PYTHIAn{}. Predictions, colour coding and bands as in Fig.~\ref{fig:ZZ-ZZmass}.}
\label{fig:ZZ-pTj}
\end{figure} 

\subsection{$ZZ$ production}
\label{sec:nlopsresultsZZ}

In Figs.~\ref{fig:ZZ-ZZmass}-\ref{fig:ZZ-pTj} we present numerical results at 
NLO, LHE level and NLO matched to \PYTHIAn{} (NLO+PS) for gluon-induced $ZZ$ production, showing the full result as well as the signal, background and interference contributions separately.

In Fig.~\ref{fig:ZZ-ZZmass} the invariant mass of the four-lepton system is shown. 
The Higgs-mediated signal shows the resonance peak at the Higgs boson mass together with the well-known significant offshell tail. The background clearly exhibits a single-resonant peak at $m_{4\ell}=m_Z$~\footnote{Interestingly we note that this single-resonant peak is missing at LO due to vanishing of the corresponding amplitudes induced by the triangle anomaly.} and increases significantly for $m_{4\ell}> 2m_Z$, where both intermediate $Z$ bosons can become onshell.
 In this region  the interference also starts to become relevant. As a consequence of the very inclusive phase-space cuts employed in our numerical analysis, both the signal and the interference reach about 10\% of the full result at large $m_{4\ell} \approx 2m_t$, with the interference being destructive. It is well known that the interference provides an even larger destructive contribution at higher values of $m_{4\ell}$, which are however beyond the validity of the $1/mt$ expansion used in our calculation. 
The $m_{4\ell}$ observable is inclusive in QCD radiation and consequently 
parton-shower corrections are marginal for all contributions (individually and in their sum). In fact, for all production modes the fixed-order NLO prediction agrees at the percent level with both the LHE level prediction and the fully showered prediction. Scale uncertainties at the fully showered level are approximately 20\%. At small invariant masses ($m_{4\ell}<150$ GeV) the interference becomes very small and consequently Monte Carlo statistics deteriorate quickly in this regime.

Fig.~\ref{fig:ZZ-HTtot} shows the distribution in 
\begin{align}
H_T=\sum\limits_{i \in \{ \ell,\nu,j\}} p_{\rm T,i} \,,
\end{align}
where the sum over the transverse momenta considers all leptons and reconstructed jets. In this distribution the signal peaks at $H_T=m_H$, while the background peaks at $H_T=2 m_Z$. For small $H_T$ parton-shower corrections are mostly driven by the first radiation already present at the LHE level. For the background contribution, these corrections are small, but for the signal contribution they lead to a negative correction of about 50\%.
A possible explanation is that the signal distribution is strongly peaked around  $m_H$ and therefore very sensitive to additional radiation that moves events away from the peak.  For large $H_T$, the
parton showers provide substantial positive corrections up to a factor of 2, while the scale uncertainties can be as large as 50\%.
This effect can be understood as follows. 
The upper cut on the invariant mass of the four leptons Eq.~\ref{eq:cut_m4l} also restricts $H_T < 340$ GeV at LO. However, the phase space for $H_T > 340$ GeV can be filled via additional QCD radiation. This leads to significant NLO corrections (not shown here), as well as to
sizable  parton-shower corrections and LO-like scale uncertainties. 

\begin{figure}[tb]
\includegraphics[width=0.48\textwidth]{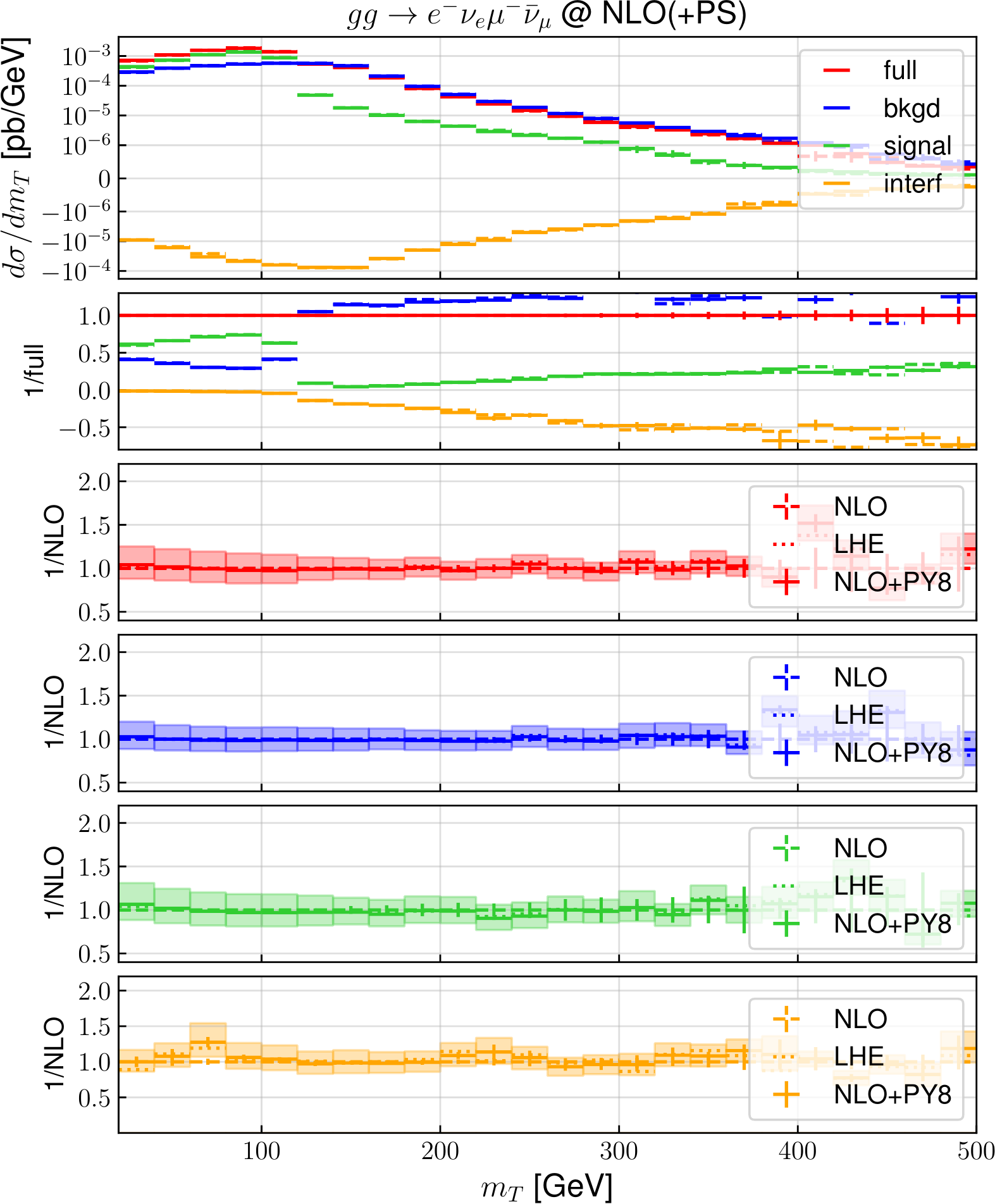}
\caption{Differential distribution in transverse mass $m_{T}$ of the four-lepton system in $gg \to e^{+} \nu_e \mu^{-} \bar\nu_{\mu} $ at NLO matched to \PYTHIAn{}. Predictions, colour coding and bands as in Fig.~\ref{fig:ZZ-ZZmass}.}
\label{fig:WW-mT}
\end{figure} 

\begin{figure}[tb]
\includegraphics[width=0.48\textwidth]{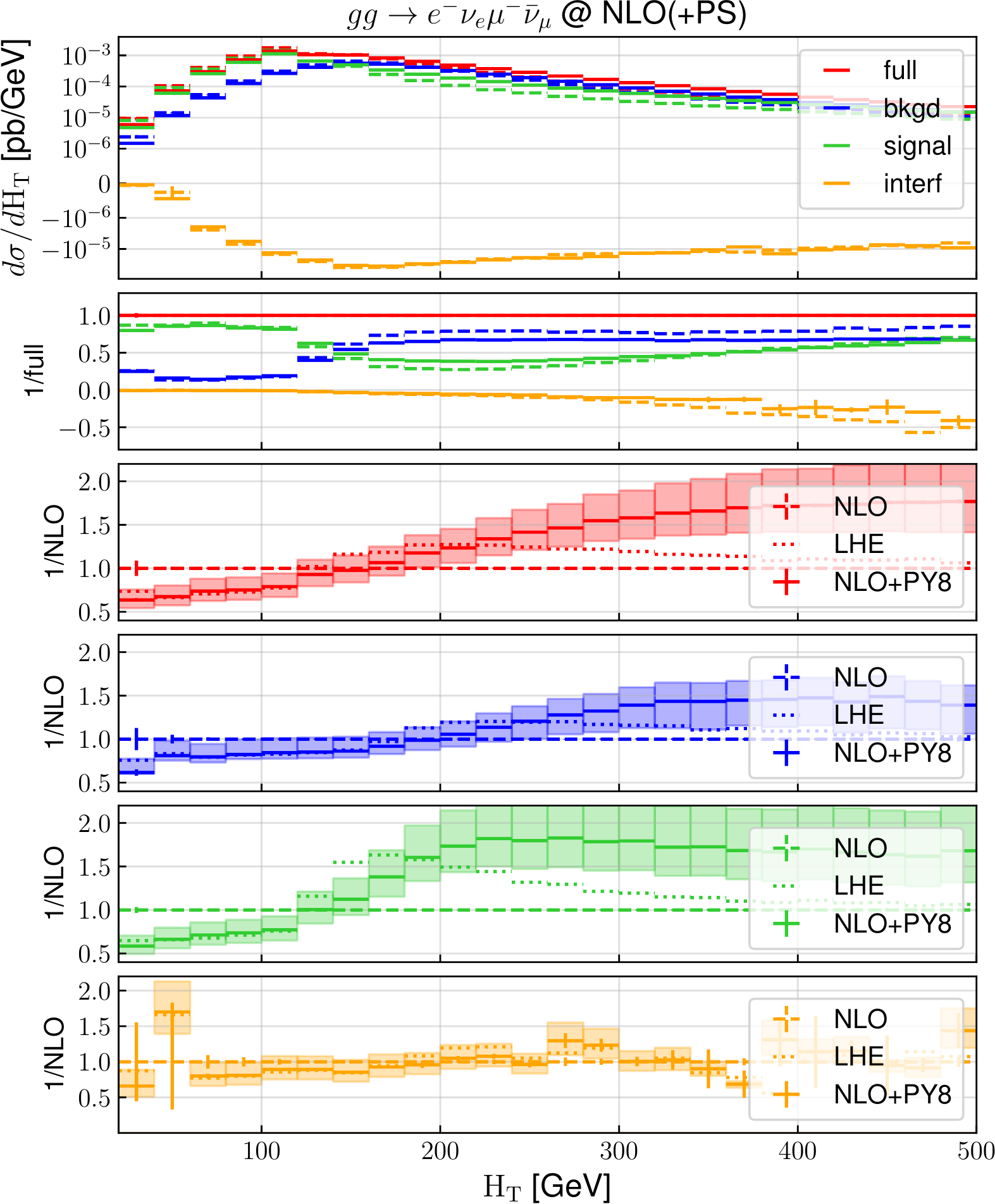}
\caption{Differential distribution in $H_{\rm T}$ in $gg \to e^{+} \nu_e \mu^{-} \bar\nu_{\mu} $ at NLO matched to \PYTHIAn{}. Predictions, colour coding and bands as in Fig.~\ref{fig:ZZ-ZZmass}.}
\label{fig:WW-HTtot}
\end{figure} 


Figs.~\ref{fig:ZZ-pTZZ} and \ref{fig:ZZ-pTj} display the transverse momentum of the four-lepton system and of the hardest jet respectively. For the latter no  lower cut on the jet transverse-momentum is applied. The two distributions are identical at fixed-order (they only differ in the first bin which for $p_{\rm T,4\ell}$ includes the Born and virtual contributions proportional to $\delta(p_{\rm T,4\ell})$). The fully showered predictions include a Sudakov suppression which can clearly be seen at the lower end of both the $p_{\rm T,4\ell}$  and the $p_{\rm T,j_1}$ distributions. We also observe that the parton shower changes the sign of the lowest bin in the $p_{\rm T,4\ell}$ spectrum. This can be understood as follows: the virtual contribution, proportional to $\delta(p_{\rm T,4\ell})$, always comes with an opposite sign of the corresponding real contribution.After the shower (and even after the first \POWHEG{} emission) the virtual contribution gets spread out at finite values of $p_{\rm T,4\ell}$. 
This results in a change of sign in the first bin.

Turning now to the opposite end of the spectrum, the $p_{\rm T,4\ell}$ distribution corresponds to the entire QCD recoil of the four-lepton system and for all contributions
receives large parton shower corrections in the tail, while LHE level corrections are largest at $p_{\rm T,4\ell}\approx 40-50\,\GeV$ and become small in the tail, where the Sudakov suppression fades away. As already discussed in Ref.~\cite{Alioli:2016xab}
the large parton-shower corrections can be explained by the fact that, by adding further radiation, the shower
increases the transverse momentum of the colour-neutral four lepton system, which has to recoil against the sum of all emitted particles.
On the contrary, in the tail of $p_{\rm T,j_1}$ no such enhancement of the corrections due to the parton shower is observed. In fact, by construction the shower emissions are subdominant with respect to the leading jet
and on average are separated enough not to be clustered with it. 
With respect to the LHE level we observe small and negative parton-shower corrections, being compatible within scale uncertainties.

\subsection{$W^+W^-$ production}

\begin{figure}[tb]
\includegraphics[width=0.48\textwidth]{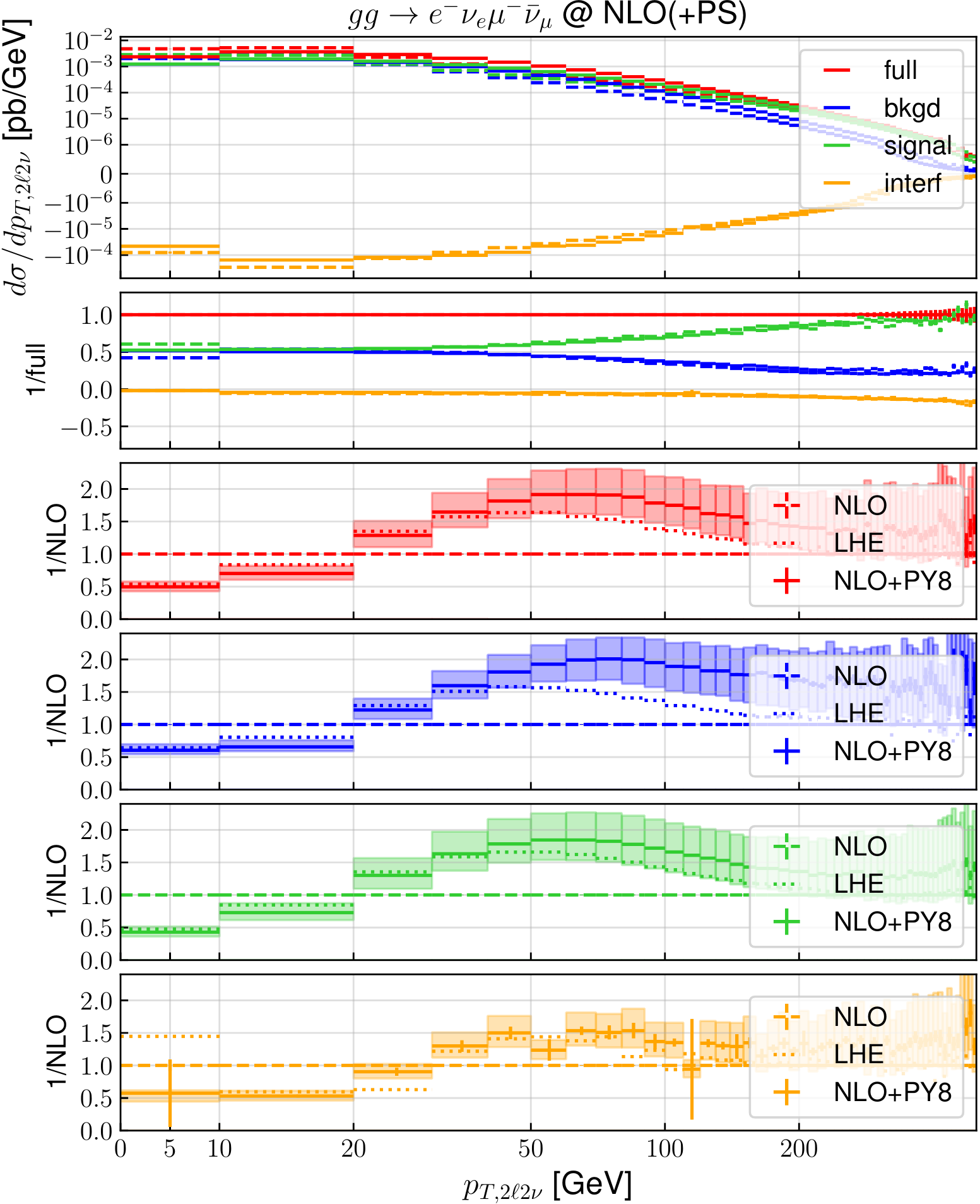}
\caption{Differential distribution in the transverse momentum of the four-lepton system $p_{\rm T,2\ell 2\nu}$ in $gg \to e^{+} \nu_e \mu^{-} \bar\nu_{\mu} $ at NLO matched to \PYTHIAn{}. Predictions, colour coding and bands as in Fig.~\ref{fig:ZZ-ZZmass}.}
\label{fig:WW-pTWW}
\end{figure}

\begin{figure}[tb]
\includegraphics[width=0.48\textwidth]{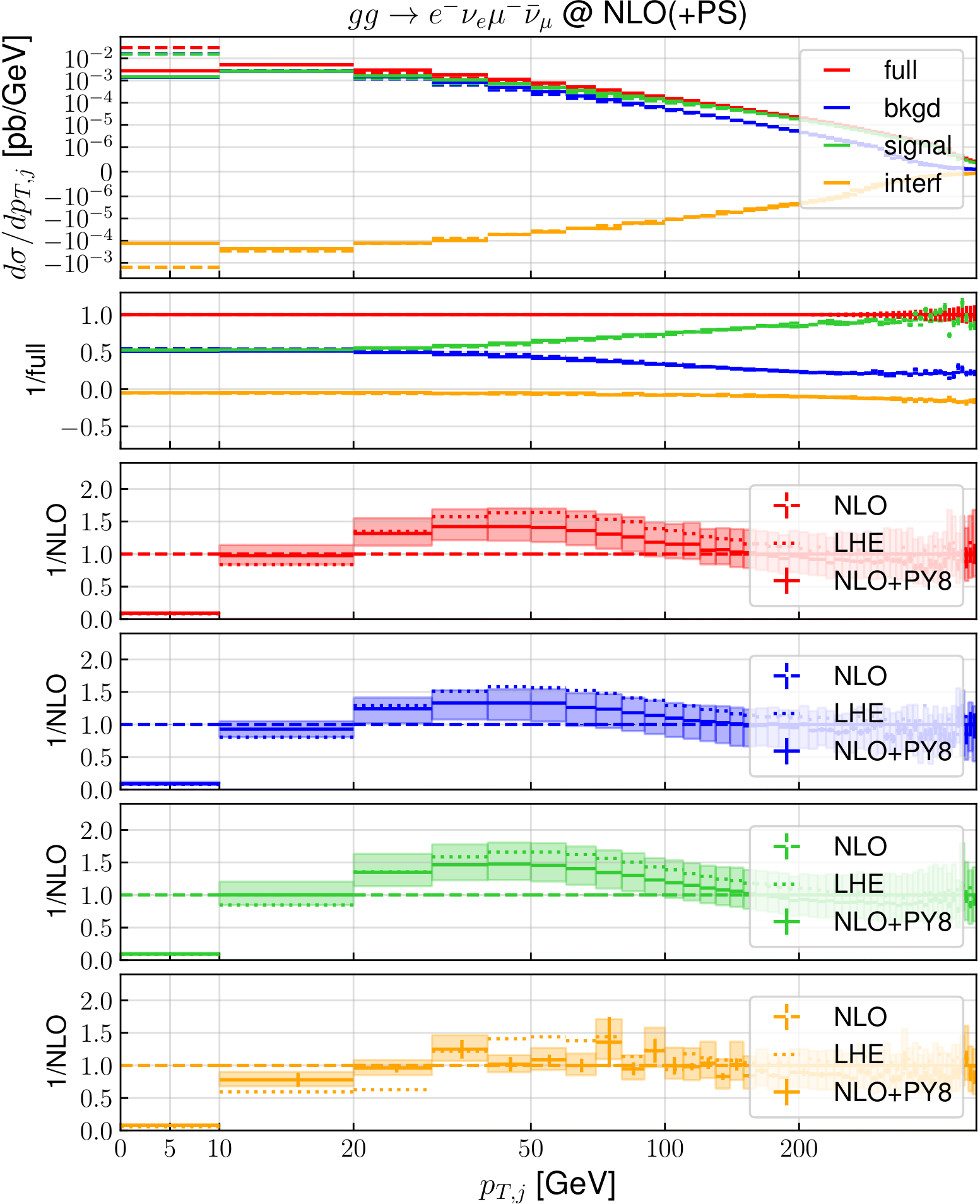}
\caption{Differential distribution in the transverse momentum of the hardest jet $p_{\rm T,j_1}$ in $gg \to e^{+} \nu_e \mu^{-} \bar\nu_{\mu} $ at NLO matched to \PYTHIAn{}. Predictions, colour coding and bands as in Fig.~\ref{fig:ZZ-ZZmass}.}
\label{fig:WW-pTj}
\end{figure}

In Figs.~\ref{fig:WW-mT}-\ref{fig:WW-pTj} we present numerical results at 
NLO, LHE level and NLO matched to \PYTHIAn{} for gluon-induced $W^+W^-$ production, showing again the signal, background, interference, and full results. 

In contrast to the corresponding results for $ZZ$ production, here we consider the 
distribution in the transverse mass $m_{T}$ of the four-lepton system, as defined in Eq. \ref{eq:mt}, 
 instead of the invariant mass of the colour-singlet system. This is shown in Fig. \ref{fig:WW-mT}. As for the invariant mass in $ZZ$ production,
the impact of the parton-shower corrections on the transverse mass in $W^+W^-$ production  is marginal,
as expected from its inclusive (with respect to QCD radiation) nature.
It is noteworthy that the interference becomes very large at high $m_T$ and 
eventually contributes beyond $-50\%$ for $m_{T}>300\,\GeV$. However, also for the interference alone parton-shower corrections are marginal for the entire  $m_T$ range considered.

A similarly strong enhancement of the interference can also be observed at large 
$H_{\rm T}$, as shown in Fig.~\ref{fig:WW-HTtot}. In the tail of this observable, parton-shower corrections are again sizable. However in contrast to $ZZ$ production, here no 
upper boundary on the four-lepton invariant mass is applied and the parton-shower
corrections level off for large $H_{\rm T}$ at around $50\%$ for the background and $70\%$ for the full.

We finally consider the QCD recoil for $W^+W^-$ production in Fig.~\ref{fig:WW-pTWW}
and the transverse-momentum distribution of the hardest jet in Fig.~\ref{fig:WW-pTj}. We observe similar behaviour as for $ZZ$ production: the anticipated Sudakov suppression at the low end of both spectra, very large parton-shower corrections in the tail of $p_{\rm T,2\ell 2\nu}$, and mild corrections in the entire  $p_{\rm T,j_1}$ spectrum.

\subsection{Shower recoil scheme}
\label{sec:recoil}

\begin{figure}[tb]
\includegraphics[width=0.48\textwidth]{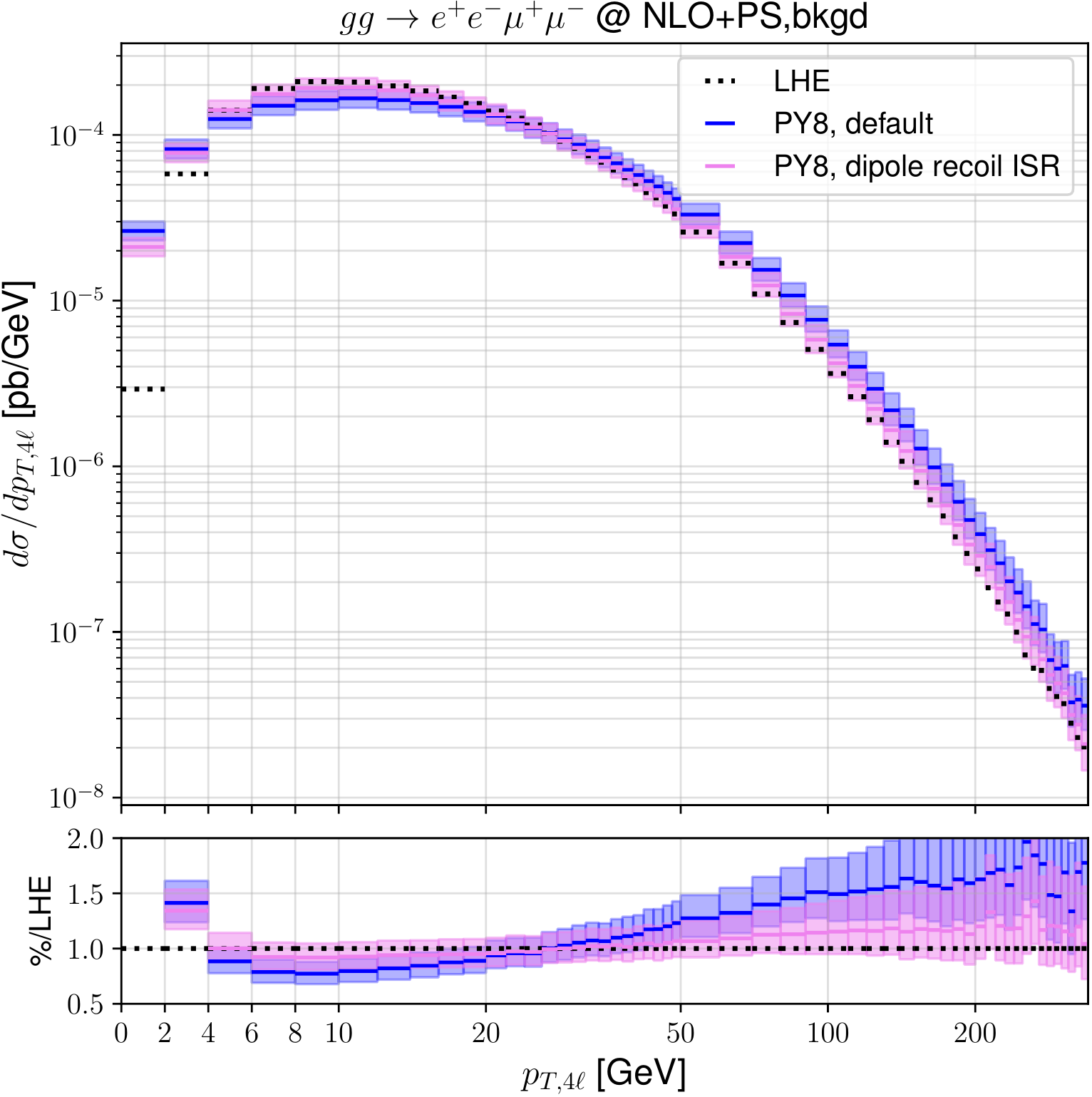}
\caption{Differential distribution in transverse momentum of the four charged leptons in  $gg \to e^+ e^- \mu^+ \mu^- $ for the background contribution at the LHE level~(black), and at the particle level, using the default \PYTHIAn{} recoil~(blue) and the fully local  dipole recoil~(violet).
  In the lower panel the ratio with respect to the LHE level is shown.
}
\label{fig:py8recoil}
\end{figure}

As discussed in Sec.~\ref{sec:matching} \PYTHIAn{} implements two alternative shower recoil schemes: the default scheme in which the transverse momentum imbalance after an initial-final dipole emission is democratically distributed among all final-state particles, including the four lepton system,  and a fully local scheme, in which the recoil is entirely absorbed by the coloured spectator.\footnote{This is activated by the \PYTHIAn{} setting \texttt{SpaceShower::dipoleRecoil =  on}.} In Fig.~\ref{fig:py8recoil} we compare these two schemes considering the transverse momentum of the four lepton system in the background contribution to $ZZ$ production.

As already anticipated in Sec.~\ref{sec:nlopsresultsZZ}
the default recoil scheme leads to a very hard spectrum in the tail (with a 50\% increase with respect to the LHE distribution around 100~GeV). Conversely the dipole scheme remains close to the LHE level at large $p_{T,4\ell}$.
For small values of $p_{T,4\ell}$, the dominant contribution should arise from several (soft) emissions whose total transverse momentum sum up to zero. However, in the dipole scheme, the transverse momentum recoil for ISR is not always absorbed by the final-state colour singlet. This explains why for very small values of $p_{T,4\ell}$ the local recoil leads to a significantly smaller cross section compared to the default scheme.
Thus, the default scheme yields a better description of the logarithmically enhanced region, while it also overpopulates the hard region of the spectrum.

A detailed discussion of the logarithmic accuracy of the parton shower goes beyond the purposes of this article and
 the choice of the recoil scheme has important implications at higher logarithmic orders~\cite{Nagy:2009vg,Dasgupta:2018nvj,Bewick:2019rbu,Dasgupta:2020fwr,Forshaw:2020wrq,Hamilton:2020rcu}. However, since the choice of the recoil scheme only affects our predictions beyond the claimed accuracy, a comparison of the two options available in \PYTHIAn{} should help assess the size of the total theoretical uncertainty.

\subsection{Effect of $qg$  and $q\bar q$ channels}
\label{sec:qgchannel}

\begin{figure}[tb]
\includegraphics[width=0.48\textwidth]{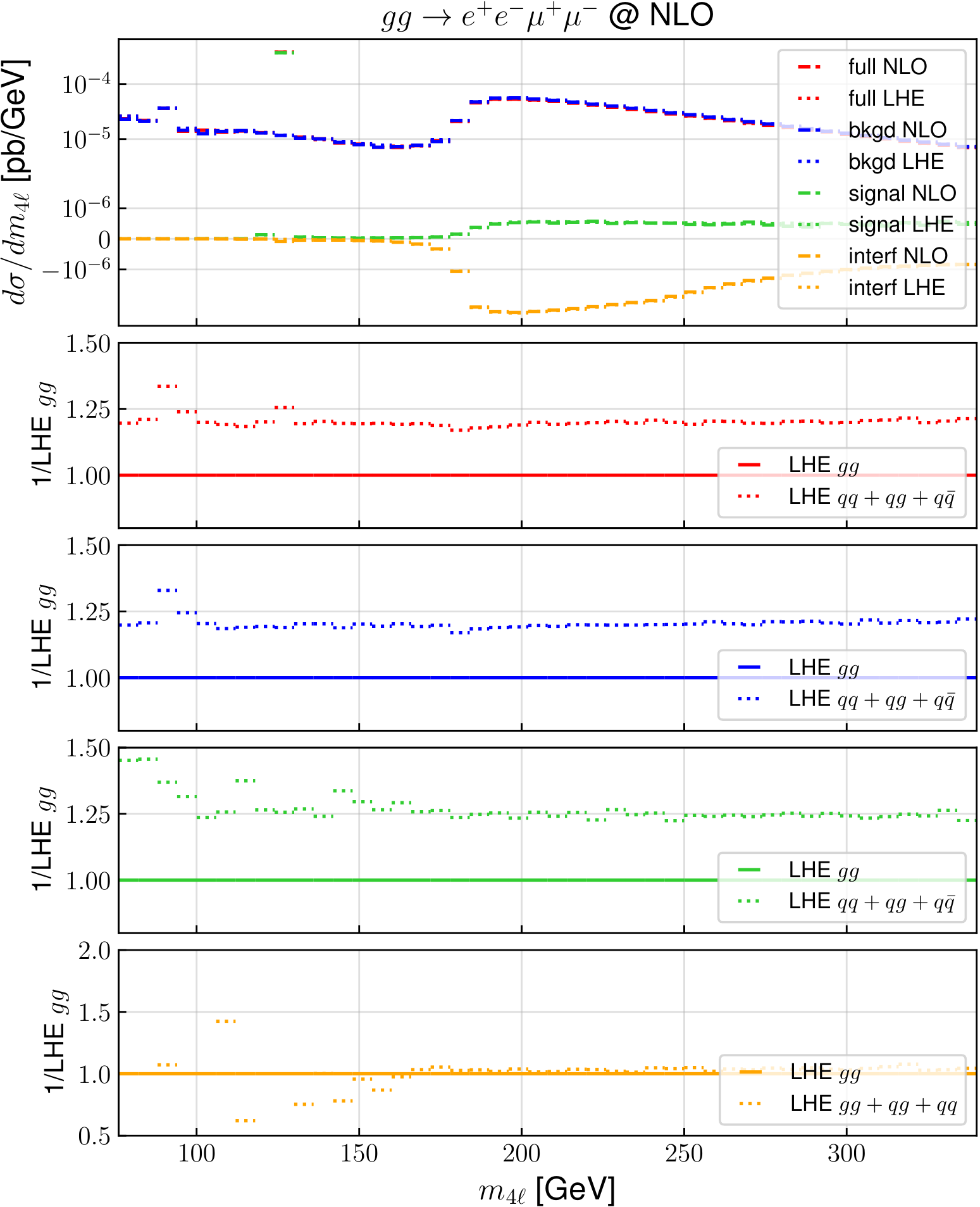}
\caption{Differential distribution in invariant mass $m_{4\ell}$ of the four-lepton system in $gg \to e^{+} e^{-} \mu^{+} \mu^{-}$ at NLO and LHE level. Colour coding of the different production modes as in Fig. \ref{fig:ZZ-ZZmass-split}. The lower ratio plots show the full LHE contribution including all partonic channels ($gg+qg+qq$) over only the $gg$ channel contribution for the different production modes.}

\label{fig:ZZ-ZZmass-split}
\end{figure}

\begin{figure}[tb]
\includegraphics[width=0.48\textwidth]{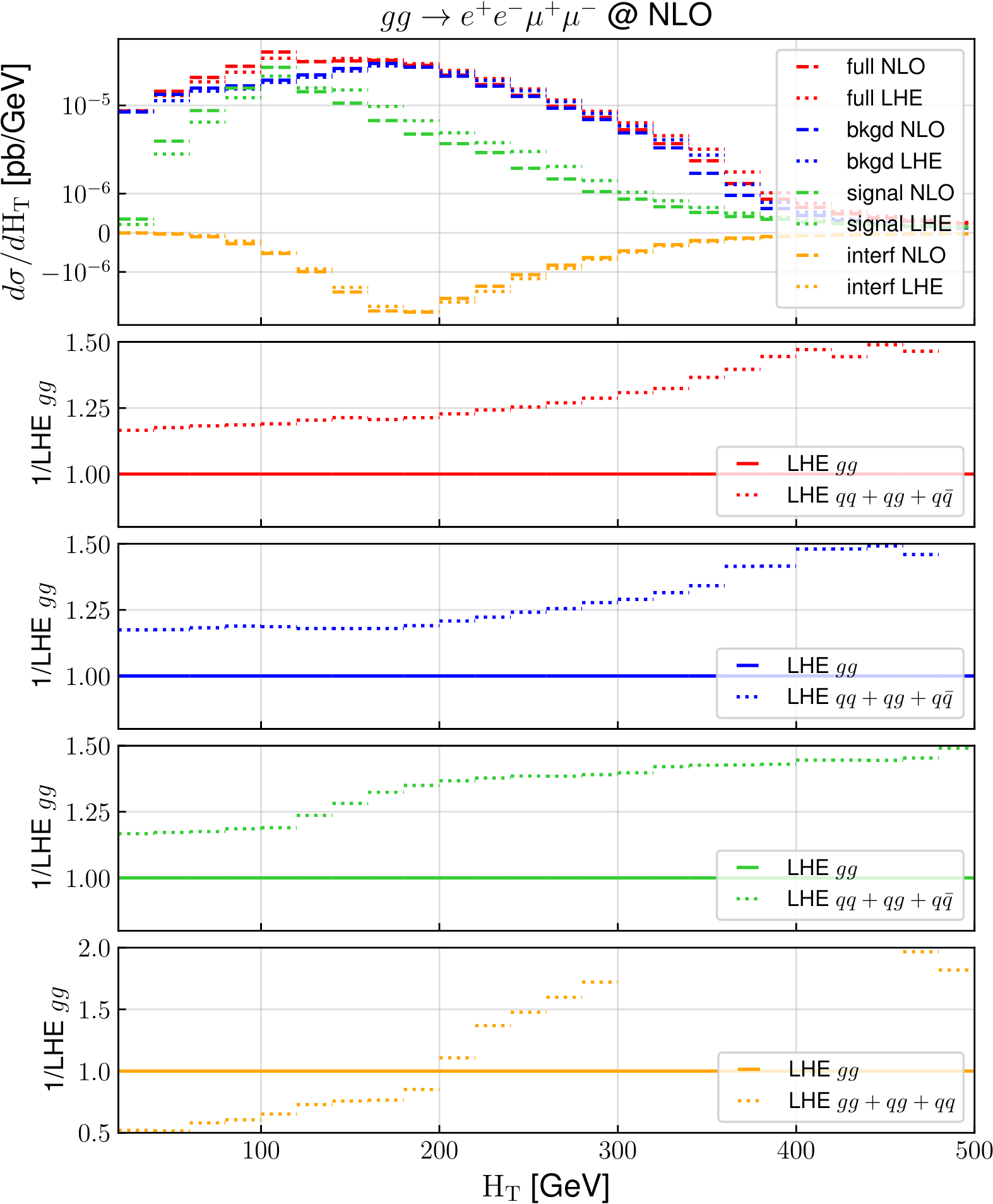}
\caption{Differential distribution in $H_{\rm T}$ in $gg \to e^+ e^- \mu^+ \mu^- $  at NLO and LHE level.  Predictions, colour coding and bands as in Fig.~\ref{fig:ZZ-ZZmass-split}.}
\label{fig:ZZ-HTtot-split}
\end{figure}

In contrast to the calculations in Ref.~\cite{Caola:2016trd,Alioli:2016xab}, in this study we do include the $qg$ and $q\bar q$ induced channels contributing to the real radiation at NLO.~\footnote{These channels were also considered in the fixed-order NLO study of \cite{Grazzini:2018owa,Grazzini:2020stb}.} Here we would like to explicitly highlight the impact of these production channels. To this end in Figs.~\ref{fig:ZZ-ZZmass-split}-\ref{fig:ZZ-HTtot-split} we illustrate at the LHE level the impact of the $qg$ and $q\bar q$ channels with respect to only the $gg$ channels for the different production modes, considering the $m_{4\ell}$ and $H_{\rm T}$ distributions in $ZZ$ production. We find very similar results also for the $W^+W^-$ production mode. In the $m_{4\ell}$ distribution the impact of the $qg/q\bar q$ channels is rather flat and about $25\%$ for all production modes. For $H_{\rm T}$ it is increasing with increasing $H_{\rm T}$ and reaches 
up to $50\%$ in the considered range. Clearly, any precision analysis of $gg$-induced four-lepton production should include these additional partonic channels opening up at NLO.

\section{Conclusions and Outlook}
\label{sec:conclusions_outlook}

Gluon-induced four-lepton production offers a unique laboratory for the measurements of offshell Higgs bosons. At the same time precision studies of diboson processes and
corresponding background estimates in new physics searches are becoming sensitive to the accuracy of the modelling of the gluon-induced production modes.
Having this in mind, in this paper we presented an implementation of the loop-induced processes $gg \to ZZ$ and $gg \to W^+W^-$ including offshell leptonic decays and non-resonant contributions at NLO matched to the \PYTHIAn{} parton shower event generator. 
 We consistently include the continuum background contribution, the Higgs-mediated signal, and their interference. All of these are loop-induced processes and therefore their implementation in a fully-exclusive NLO event generator matched to parton showers poses a significant technical challenge.

 In inclusive observables, such as the four-lepton\linebreak invariant-mass distribution in $ZZ$ production, the \linebreak parton-shower corrections are found to be marginal, while in more exclusive observables like the recoil of the four-lepton system they can become substantial. For the latter we highlighted the importance of the parton-shower recoil scheme. Furthermore we investigated the relevance of the $qg/q\bar q$ induced production channels, which partly overlap with the higher-order corrections to quark-induced diboson production.

In our calculation all ingredients have been treated exact at the NLO level apart from the massive amplitudes contributing to the two-loop virtuals, which are incorporated via approximations.  
Exact results for the latter have become available very recently and could be incorporated in an updated version of the \ggfl generator presented here. Moreover, the generator will be made publicly available in the \RES framework.

\section*{Acknowledgments}
We are grateful to Fabrizio Caola for useful discussions during the preliminary stages of this work. We also thank Paolo Nason for discussing the modifications to \RES{} necessary for implementing the processes described in this paper.
J.L. is supported by the Science and Technology Research Council (STFC) under the Consolidated Grant ST/T00102X/1 and the STFC 
Ernest Rutherford Fellowship ST/S005048/1.
The work of S.A. is
supported by the ERC Starting Grant REINVENT-714788. He also acknowledges
funding from Fondazione Cariplo and Regione Lombardia, grant 2017-2070 and by
the Italian MUR through the FARE grant R18ZRBEAFC.
S.F.R.'s work was
supported by the European Research Council (ERC) under the European
Union’s Horizon 2020 research and innovation programme (grant
agreement No. 788223, PanScales) and by the UK Science and Technology
Facilities Council (grant number ST/P001246/1).

\appendix

\section{Modifications of the \RES framework}
\label{sec:mod}
In this section we outline the modifications we implemented to the \RES framework in order to be able to deal with a loop-induced process and with non-positive defined LO processes. These modifications are available in the \texttt{gg4l} process folder and will be incorporated in a future release of the \RES.

As already discussed in Sec.~\ref{sec:computational_setup}, we discard configurations where the one-loop real amplitude becomes unstable, on the ground that this happens only when the $p_T$ of the radiated parton is very small and thus, once we include the subtraction terms, we are left only with negligible power corrections. To do so, setting to zero the amplitude computed in {\tt setreal} is not sufficient and we have modified the subroutine {\tt btildereal} to ensure that the subtraction terms are included only when a non-zero real amplitude is found.

For the real corrections, we have the possibility to select only the $gg\to VV g$ channel. By doing so, we are excluding the collinear divergent term $q(\bar{q})g \to VV q(\bar{q})$. Thus, we have modified the subroutine {\tt btildecoll} to include the integrated $q(\bar{q}) \to g q(\bar{q})$ collinear remnant only when the real $q(\bar{q})g \to VV q(\bar{q})$ contribution is considered.

The \texttt{NNPDF30\_nlo\_as\_0118} PDF set includes non-perturbative corrections which becomes sizeable for scales smaller than $m_b=4.5$~GeV. 
This may cause problems in estimating the upper bound for the strong coupling when generating the radiation according to the method detailed in Ref~\cite{Alioli:2010xd}. In essence, the upper bound is computed by using the LO running of $\alpha_s$ and a suitable choice of the infrared cutoff $\Lambda_{\rm rad}$, which controls the magnitude of the running, so that
\begin{equation}
\alpha_s^{\rm upb} (p_T)=\frac{1}{2 b_0 \log\frac{p_T}{\Lambda_{\rm rad}}} \ge \alpha^{\rm cmw}_s(p_T),
\end{equation}
with $b_0=\frac{33-2\times 5}{12\pi}$, and ``cmw'' denoting the Catani-Marchesini-Webber prescription for the running coupling~\cite{Catani:1990rr}.
  We have modified the appropriate subroutine ({\tt init\_rad\_lambda}) in such a way one spans over scales smaller than the bottom threshold to find the appropriate value of $\Lambda_{\rm rad}$.

  When performing event generation, the subroutines in \RES{} implicitly assume that both  the Born and the real squared amplitudes are positive. This is not the case when we consider the interference contribution alone, which can also be negative. Thus, we have modified the subroutines {\tt gen\_rad\_isr}, {\tt pick\_random}  and {\tt do\_maxrat} to work with absolute values. Furthermore, away from the singular limits, Born and real amplitudes can have opposite signs. When this happens, we always assume that the real contribution is nonsingular and we do not apply any \POWHEG{} Sudakov suppression to it. Therefore we move these nonsingular contributions into the remnants by means of a modified {\tt bornzerodamp} subroutine.

\bibliographystyle{elsarticle-num}

\end{document}